\documentclass[12pt,a4paper]{article}
\usepackage[dvipdfmx]{graphicx}
\usepackage{color}
\usepackage{citesort}
\usepackage{latexsym}
\usepackage{amsmath}
\usepackage{amssymb}
\usepackage{url}

\newcommand{\ie}{\textit{i.e.}}
\newcommand{\eg}{\textit{e.g.}}
\newcommand{\eref}[1]{Eq.~\eqref{eq:#1}}
\newcommand{\figref}[1]{Fig.~\ref{fig:#1}}

\newcommand{\meank}{\langle k \rangle}

\newcommand{\rhosi}{\rho_{{\rm S}, i}}
\newcommand{\rhoii}{\rho_{{\rm I}, i}}
\newcommand{\rhosj}{\rho_{{\rm S}, j}}
\newcommand{\rhoij}{\rho_{{\rm I}, j}}
\newcommand{\rhoi}{\rho_{i}}
\newcommand{\Ds}{D_{\rm S}}
\newcommand{\Di}{D_{\rm I}}
\newcommand{\rhosk}{\rho_{{\rm S}, k}}
\newcommand{\rhoik}{\rho_{{\rm I}, k}}
\newcommand{\rhoskp}{\rho_{{\rm S}, k^{\prime}}}
\newcommand{\rhoikp}{\rho_{{\rm I}, k^{\prime}}}
\newcommand{\rhok}{\rho_{k}}
\newcommand{\rhoS}{\rho_{\rm S}}
\newcommand{\rhoI}{\rho_{\rm I}}
\newcommand{\betac}{\beta_{\rm c}}
\newcommand{\fia}{f_{\rm I}^*}

\begin{document}
\thispagestyle{empty}

\title{Sufficient conditions of endemic threshold on metapopulation networks}

\author{Taro Takaguchi$^{1,2,\dagger}$ and Renaud Lambiotte$^{3,\ddagger}$\\
\\
\\
${}^{1}$ 
National Institute of Informatics,\\
2-1-2 Hitotsubashi, Chiyoda-ku, Tokyo 101-8430, Japan
\\
\\
${}^{2}$
JST, ERATO, Kawarabayashi Large Graph Project,\\
2-1-2 Hitotsubashi, Chiyoda-ku, Tokyo 101-8430, Japan
\\
\\
${}^{3}$
Department of Mathematics and naXys, University of Namur,\\
5000 Namur, Belgium
\\
\\ 
${}^\dagger$t\_takaguchi@nii.ac.jp\\
${}^\ddagger$renaud.lambiotte@unamur.be\\
}

\setlength{\baselineskip}{0.77cm}

\maketitle

\noindent
\copyright 2015, Elsevier. Licensed under the Creative Commons Attribution-NonCommercial-NoDerivatives 4.0 International \url{http://creativecommons.org/licenses/by-nc-nd/4.0/}
\newpage

\begin{abstract}
\setlength{\baselineskip}{0.77cm}
In this paper, we focus on susceptible-infected-susceptible dynamics on metapopulation networks, where nodes represent subpopulations, and where agents diffuse and interact.
Recent studies suggest that heterogeneous network structure between elements plays an important role in determining the threshold of infection rate at the onset of epidemics, a fundamental quantity governing the epidemic dynamics.
We consider the general case in which the infection rate at each node depends on its population size, as shown in recent empirical observations. 
We first prove that a sufficient condition for the endemic threshold (\ie, its upper bound), previously derived based on a mean-field approximation of network structure, also holds true for arbitrary networks.
We also derive an improved condition showing that networks with the rich-club property (\ie, high connectivity between nodes with a large number of links) are more prone to disease spreading.
The dependency of infection rate on population size introduces a considerable difference between this upper bound and estimates based on mean-field approximations, even when degree--degree correlations are considered.
We verify the theoretical results with numerical simulations.
\end{abstract}

\begin{center}
\end{center}
\newpage

\section{Introduction}

Spreading of infectious diseases is one of the major threats to human society and understanding their dynamics is crucial to control them.
When modeling the dynamics of infectious diseases in large-scale populations, it is often useful to adopt a metapopulation approach~ \cite{Anderson1991,Hanski2000,Colizza2007a}. The detailed, most of the time unknown, network of contacts between individuals is then replaced by a coarse-grained network, where nodes represent regions in which individuals interact, \eg, cities, and where links correspond to flows of individuals between the nodes. The computational and mathematical analysis of metapopulation models aims at identifying how network topology and model parameters affect the global epidemics. In a majority of realistic scenarios, nodes of metapopulation networks tend to exhibit a strong heterogeneity in their size and their degree. For instance, in the case of network of cities, it is well known that the size of cities is distributed according to a power-law~\cite{Zipf1949,Gabaix1999}, and that their total flow is well approximated by their size~\cite{Carrothers1956}. Recent research has mainly focused on the impact of the degree distribution on epidemic spreading~\cite{Colizza2007a,Saldana2008,Colizza2008,Juher2009,Saldana2010,Juher2014}, but the effect of their size distribution remains, to our knowledge, poorly known. 

Directly related to this problem, the dependency of the contact rate between individuals on the size of population is a central theme in mathematical epidemiology~\cite{Smith2009}. To date, most studies have used terms that are either population-dependent, where the contact rate varies linearly with population size, or population-independent, where the rate is constant. In the literature, directly transmitted diseases such as measles, and foot and mouth disease, tend to be described by population-dependent models~\cite{Earn2000,Tildesley2006}, whereas sexually transmitted diseases such as HIV are usually considered population-independent~\cite{May1987}. More importantly, studies have shown that models parameterized with different transmission terms can predict very different quantitative and qualitative infection dynamics~\cite{McCallum2001}. These observations suggest an intermediate scenario, where the scaling of contact rate on population size takes the general form $N^{\gamma}$ with $0 \leq \gamma \leq 1$. In this direction, empirical measures on rodents have shown that the standard modeling parameters, $\gamma=0$ and $\gamma=1$ can be significantly rejected, and that the observed scaling exponent lies between these two extremes~\cite{Smith2009}. In a different context, in the urban sciences, where the size of cities is known to play a fundamental role for collective social behaviors~\cite{Bettencourt2007}, recent research based on mobile phone data has also shown that the total number of contacts grows super-linearly with city population size~\cite{Schlapfer2014}. Taken together, these findings motivate the study of metapopulation models with general contact rates, and a better understanding of the impact of city size distribution on epidemic spreading.

\section{Model}
\subsection{Definition of the reaction--diffusion process}
We consider a metapopulation network (we simply call it the network hereafter), composed of $N$ nodes with $n_p$ particles on it.
Particles move between nodes through the links and nodes serve as substrate of interaction between particles.

In this paper, we assume that particles interact with each other according to so-called susceptible-infected-susceptible (SIS) dynamics~\cite{Anderson1991}, in line with the preceding works~\cite{Colizza2007a,Saldana2008,Colizza2008,Juher2009,Saldana2010,Juher2014,Masuda2010,Iggidr2012,Lund2013}.
In the SIS dynamics, a particle takes either of two states, \ie, susceptible (S) and infected (I),
and the infection and recovery processes are described as follows:   
\begin{align}
\begin{cases}
{\rm S} + {\rm I} \to {\rm I} + {\rm I} & {\rm with \ rate} \ \beta_i,\\
{\rm I} \to {\rm S} & {\rm with \ rate} \ \mu .
\end{cases}
\end{align}
Here we assume that the infection rate $\beta_i$ is different for every node~$i$ ($1 \leq i \leq N$). 
Aside from the reactions, particles perform diffusion between nodes with diffusion coefficient $\Ds$ and $\Di$ for susceptible and infected particles, respectively.
In general, SIS dynamics has 
an absorbing state corresponding to the the disease-free equilibrium where there is no infected particle.
The SIS dynamics exhibits a phase transition at a threshold value $\betac$ such that in the case of $\beta > \betac$ the endemic equilibrium emerges and there is a finite fraction of infected particles at the stationary state ~\cite{Dorogovtsev2008}.
The estimation of $\betac$ for arbitrary metapopulation networks is the main concern of the present study.

We assume that interaction between particles within a node and diffusion of particles between nodes occur simultaneously~\cite{Saldana2008},
instead of the two-step process of reaction and diffusion investigated in \cite{Colizza2007a}.
In the continuous time limit, the average behavior of the dynamics can be described by the deterministic reaction-diffusion equation on an arbitrary network~\cite{Masuda2010} as
\begin{align}
\partial_t \rhosi &= \mu \rhoii - \frac{\beta_i}{\rhoi} \rhosi \rhoii - \Ds \sum_{j} L_{ij} \rhosj,\label{eq:rhosi}\\
\partial_t \rhoii &= -\mu \rhoii + \frac{\beta_i}{\rhoi} \rhosi \rhoii - \Di \sum_{j} L_{ij} \rhoij,\label{eq:rhoii}
\end{align}
where $\rhosi$ and $\rhoii$ represent the number of susceptible and infected particles at node~$i$, respectively.
The total number of particles at node~$i$, denoted by $\rhoi$, is equal to $\rhosi + \rhoii$.
Matrix $L$ denotes the random-walk Laplacian whose elements are given by
\begin{align}
L_{ij} =
\begin{cases}
1 & (i = j),\\
-A_{ij} / k_j & (i \neq j),
\end{cases}
\end{align}
where $A_{ij}$ is the element of adjacency matrix $A$ and equal to unity if nodes $i$ and $j$ are adjacent and equal to zero otherwise.
Degree $k_i$ of node~$i$ represents the number of links connected to the node.
In this paper, we assume that the network is connected (\ie, consists of a single connected component) via undirected and unweighted links and is simple.
In other words, $A$ is a symmetric matrix, with diagonal elements $A_{ii} = 0$ ($1 \leq i \leq N$). 
It should be noted that the total number of particles is conserved throughout the process. 
The average number of particles on a node is denoted by $\rho \equiv n_p / N$.
 
The infection rate at node~$i$ is set to depend on $\rhoi$ as a power-law scaling given by $\beta_i \equiv \beta \rhoi^{\gamma}$ with $\gamma \geq 0$. This scaling, motivated by recent empirical findings~\cite{Smith2009,Schlapfer2014} described in the introduction, was also considered in Refs.~\cite{Anderson1982,McCallum2001,Juher2014}.
A special case $\gamma = 0$ corresponds to the density-dependent SIS dynamics \cite{Colizza2007a} and $\gamma = 1$ the so-called mass interaction which was considered in Refs.~\cite{Colizza2007a,Masuda2010}. 
Instead of considering a binary setting where $\gamma \sim 1$ for nodes with a small population and $\gamma \sim 0$ for nodes with a large population~\cite{Lund2013}, we assume the same $\gamma$ for all nodes such that infection rate continuously increases with population size.

\subsection{Heterogeneous mean-field approximation}\label{sec:hmf}
In this section, we briefly describe the known results about the dynamics under the heterogeneous mean-field (HMF) approximation~\cite{Saldana2008,Saldana2010}.
Under the HMF approximation, the nodes with the same degree are assumed to behave in the same way on average.
To be more precise, equations~\eqref{eq:rhosi} and \eqref{eq:rhoii} are reduced to
\begin{align}
\partial_t \rhosk &= \mu \rhoik - \beta {\rhok}^{\gamma-1} \rhosk \rhoik - \Ds \rhosk + \Ds k \sum_{k^{\prime}} P(k^\prime|k) \frac{1}{k^\prime} \rhoskp,\label{eq:rhosk_mean-field}\\
\partial_t \rhoik &= -\mu \rhoik + \beta {\rhok}^{\gamma-1} \rhosk \rhoik  - \Di \rhoik + \Di k \sum_{k^{\prime}} P(k^\prime|k) \frac{1}{k^\prime} \rhoikp.
\label{eq:rhoik_mean-field}
\end{align}
In uncorrelated networks, $P(k^{\prime} | k) = k^{\prime} P(k^{\prime}) / \meank$ holds true (where $\meank$ is the average degree of the network)
and the equations are simplified as
\begin{align}
\partial_t \rhosk &= \mu \rhoik - \beta {\rhok}^{\gamma-1}\rhosk \rhoik - \Ds \left( \rhosk - \frac{k}{\langle k \rangle} \rhoS \right),\label{eq:rhosk}\\
\partial_t \rhoik &= -\mu \rhoik + \beta {\rhok}^{\gamma-1} \rhosk \rhoik  - \Di \left( \rhoik - \frac{k}{\langle k \rangle} \rhoI \right), \label{eq:rhoik}
\end{align}
where $\rhoS \equiv \sum_k P(k) \rhosk$ and $\rhoI \equiv \sum_k P(k) \rhoik$.

We are interested in the condition of parameters for realizing the endemic equilibrium in the long-term limit (\ie, $t \to \infty$ and $\partial_t \rhosk, \partial_t \rhoik \to 0$).
In this limit, if $\Ds, \Di > 0$, $\rhok$ converges to the stationary distribution $\rhok^*$ regardless of the other parameters.
Equations~\eqref{eq:rhosk} and \eqref{eq:rhoik} have a fixed point $(\rhosk^*, \rhoik^*) = (k \rho / \meank, 0)$ which corresponds to the disease-free equilibrium.
The endemic equilibrium arises when the disease-free equilibrium is unstable.
The linear stability of the disease-free equilibrium is determined by the Jacobian matrix at the fixed point:
\begin{equation}
J_{\rm HMF} =
\left(
\begin{array}{cc}
J^{(1)}_{\rm HMF} & J^{(3)}_{\rm HMF} \\
O & J^{(2)}_{\rm HMF}\\
\end{array}
\right),
\label{eq:J_hmf}
\end{equation}
where $J^{(i)}_{\rm HMF}$ $(i=1,2,3)$ are $k_{\max} \times k_{\max}$ matrices and $O$ is the $k_{\max} \times k_{\max}$ zero matrix, where $k_{\max}$ is the maximum $k_i$ in the network.
The elements of $J^{(1)}_{\rm HMF}$ and $J^{(2)}_{\rm HMF}$ are given by
\begin{align}
\left( J^{(1)}_{\rm HMF} \right)_{k k^{\prime}}
&= - \Ds \left( \delta_{k, k^{\prime}} +  \frac{k}{\langle k \rangle}P(k^{\prime}) \right),\label{eq:J1_hmf}\\
\left( J^{(2)}_{\rm HMF} \right)_{k k^{\prime}} &= \left( -\mu + \beta \left( \frac{k}{\meank} \rho \right)^{\gamma} - \Di \right) \delta_{k, k^{\prime}} + \Di \frac{k}{\meank} P(k^\prime),\label{eq:J2_hmf}
\end{align}
where $\delta_{k k^\prime}$ denotes the Kronecker delta.

The disease-free equilibrium becomes unstable when the real part of the largest eigenvalue of $J_{\rm HMF}$ is positive.
As we see in Eq.~\eqref{eq:J_hmf}, the eigenvalues of $J_{\rm HMF}$ are given by the union of the eigenvalues of $J^{(1)}_{\rm HMF}$ and $J^{(2)}_{\rm HMF}$.
From \eref{J1_hmf}, $J^{(1)}_{\rm HMF}$ consists of the sum of identity matrix and a rank-$1$ matrix whose eigenvalues are zero (with degeneracy factor $k_{\max}-1$) and unity. Therefore, the eigenvalues of $J^{(1)}_{\rm HMF}$ are equal to $-\Ds$ and $-2\Ds$ and negative regardless of parameter values.
Then, the largest eigenvalue of $J_{\rm HMF}$ is given by the largest eigenvalue of $J^{(2)}_{\rm HMF}$.
The largest eigenvalue of $J^{(2)}_{\rm HMF}$ satisfies 
\begin{equation}
\max_{1 \leq k \leq k_{\max}} \left( -\mu + \beta \left( \frac{k}{\meank} \rho \right)^{\gamma} - \Di \right) \leq \lambda_{\max} ( J_{\rm HMF}^{(2)}) \leq  \max_{1 \leq k \leq k_{\max}} \left( -\mu + \beta \left( \frac{k}{\meank} \rho \right)^{\gamma} \right).
\end{equation}
Therefore, the sufficient condition of $\beta$ for positive $\lambda_{\max} ( J^{(2)}_{\rm HMF} )$ is given by
\begin{equation}
\beta > \left( \mu + \Di \right) \left( \frac{k_{\max}}{\meank} \rho \right)^{-\gamma}.
\end{equation}
and an upper bound of the endemic threshold $\betac$ is derived as
\begin{equation}
\overline{\beta}_{\rm HMF} \equiv \left( \mu + \Di \right) \left(\frac{k_{\max}}{\meank} \rho \right)^{-\gamma}
\label{eq:sufficient_hmf}
\end{equation}
This upper bound of endemic threshold is trivially determined by the most susceptible node, that is the most populated one where the infection rate is maximum in our setting.
Although we only reviewed the analysis for uncorrelated networks, the same upper bound is also proven for correlated random networks, \ie, $P(k^{\prime} | k) \neq k^{\prime} P(k^{\prime}) / \meank$ (see Theorem~4.3 in Ref.~\cite{Juher2014}).

In closing this section, it is worth to mention the relationship with the endemic threshold derived in the previous studies.
When we set $\gamma=1$ and $\Di = 1$ for the two-step reaction-diffusion modeling~\cite{Colizza2007a} and $\Di \to \infty$ for the simultaneous reaction-diffusion modeling~\cite{Masuda2010} (the same as one we adopt in this paper),
the endemic threshold $\betac$ is exactly derived as 
\begin{equation}
\betac = \frac{\meank^2}{\langle k^2 \rangle} \frac{\mu}{\rho}.
\label{eq:beta_c_k_square}
\end{equation}
This functional form of $\betac$ implies that the heterogeneity in $P(k)$ plays a crucial role in determining the critical point, as one observes in the percolation processes~\cite{Cohen2000,Callaway2000} and epidemic processes on contact networks (\ie, networks between individuals)~\cite{Pastor-Satorras2001,Pastor-Satorras2002}.
The two studies reporting this $\betac$~\cite{Colizza2007a,Masuda2010} have an assumption in common on deriving \eref{beta_c_k_square};
the diffusion of infected particles are very fast and thus $\rhoii = k_i / \meank \rhoI$ holds true.
Under this assumption, the endemic threshold $\betac$ for any value of $\gamma$ is given by \eref{beta_c_k_square}.
If this assumption does not hold true, the exact $\betac$ is no longer tractable and its upper bound is obtained as described above.

\section{Results}
\subsection{Endemic threshold for arbitrary networks}\label{sec:arbitrary}
\subsubsection{Proof of $\overline{\beta}_{\rm HMF}$ for general case}\label{sec:general_case}
We first prove that the upper bound $\overline{\beta}_{\rm HMF}$ (given by Eq.~\eqref{eq:sufficient_hmf}) also holds true for arbitrary networks (see Eq.~\eqref{eq:sufficient_arbitrary}).
The following results are essentially in parallel with previous studies~\cite{Saldana2008,Juher2009,Saldana2010,Masuda2010,Lund2013,Juher2014}.
The most important difference with these studies is that they assume a HMF approximation, except for Ref.~\cite{Masuda2010}, whereas we derive our results directly from the adjacency matrix of the network, thereby considering arbitrary networks free from this approximation.

As we saw in the derivation with the HMF approximation (see Sec.~\ref{sec:hmf}),
we consider the stability of the disease-free equilibrium $(\rhosi^*, \rhoii^*) = (k_i / \meank \rho, 0)$ $(1 \leq i \leq N)$ of the original reaction-diffusion equations (given by Eqs.~\eqref{eq:rhosi} and \eqref{eq:rhoii} with $\beta_i \equiv \beta \rhoi^\gamma$).
The linear stability of the disease-free equilibrium is determined by the Jacobian matrix at the equilibrium:
\begin{equation}
J = 
\left(
\begin{array}{cc}
J^{(1)} & J^{(3)} \\
O & J^{(2)}\\
\end{array}
\right)
\end{equation}
All matrices $J^{(1)}$, $J^{(2)}$, $J^{(3)}$, and $O$ are with size $N \times N$. 
Matrix $O$ is zero matrix.
The element of $J^{(1)}$ and $J^{(2)}$ is given by
\begin{align}
J^{(1)}_{ij} &= - \Ds L_{ij},\\
J^{(2)}_{ij} &= \left( \beta \left(\frac{k_i}{\meank}\rho \right)^{\gamma} - \mu \right) \delta_{ij} - \Di L_{ij}.
\end{align}
Because the eigenvalues of $L$ are in $0 \leq \lambda(L) \leq 2$~\cite{Chung1997}, 
the eigenvalues of $J^{(1)}$ are in $-2\Ds \leq \lambda(J^{(1)}) \leq 0$ and negative regardless of parameter values.
Therefore, to derive the condition for the endemic equilibrium, we investigate the largest eigenvalue of $J^{(2)}$, denoted by $\lambda_{\max}(J^{(2)})$, in the following.

Before moving to general cases, we note two special cases in which we can exactly derive $\betac$.
As the first case, when $\gamma=0$, the element of $J^{(2)}$ is equal to
\begin{equation}
J^{(2)}_{ij} = \left( \beta - \mu \right) \delta_{ij} - \Di L_{ij}.
\end{equation}
Therefore, $\lambda_{\max}(J^{(2)})$ is given by $(\beta - \mu)$
and it leads to $\betac = \mu$.
As the second case, when the network is regular and $k_i = k$ $(1 \leq i \leq N)$, the element of $J^{(2)}$ is equal to
\begin{equation}
J^{(2)}_{ij} = \left( \beta \rho^\gamma - \mu \right) \delta_{ij} - \Di L_{ij}.
\end{equation}
Therefore, $\lambda_{\max}(J^{(2)})$ is given by $(\beta \rho^\gamma - \mu)$
and it leads to $\betac = \mu \rho^{-\gamma}$.
It is worthy to note that the endemic threshold $\betac$ is independent of diffusion rate $\Di$ for these two cases.

Now we consider general cases with $\gamma \neq 0$ on a non-regular network. 
For the convenience for calculation, we symmetrize $J^{(2)}$ via the following similarity transformation~\cite{Masuda2010}:
\begin{equation}
\hat{J}^{(2)} \equiv {\rm diag} \left( \frac{1}{\sqrt{\mathstrut k_1}}, \frac{1}{\sqrt{\mathstrut k_2}}, \ldots, \frac{1}{\sqrt{\mathstrut k_N}} \right) J^{(2)} {\rm diag} \left(\sqrt{\mathstrut k_1}, \sqrt{\mathstrut k_2}, \ldots, \sqrt{\mathstrut k_N} \right),  
\end{equation}
where ${\rm diag}(a_1, a_2, \ldots, a_N)$ represents a diagonal matrix whose $(i,i)$-element is equal to $a_i$.
Element of $\hat{J}^{(2)}$ is given by
\begin{equation}
\hat{J}^{(2)}_{ij} = \left( \beta \left(\frac{k_i}{\meank}\rho \right)^{\gamma} - \mu - \Di \right) \delta_{ij} + \Di \frac{A_{ij}}{\sqrt{\mathstrut k_i k_j}}.
\end{equation}

When the largest eigenvalue of $\hat{J}^{(2)}$, denoted by $\lambda_{\max}(\hat{J}^{(2)})$, is positive,
the disease-free equilibrium is unstable and the endemic equilibrium arises.
As we did for the case with HMF approximation (see Sec.~\ref{sec:hmf}), we estimate a lower bound of $\lambda_{\max}(\hat{J}^{(2)})$ (that leads to an upper bound of $\betac$).
By using the Rayleigh quotient, $\lambda_{\max}(\hat{J}^{(2)})$ is given by
\begin{equation}
\lambda_{\max}(\hat{J}^{(2)}) = \max_{|x| = 1} \frac{x^\top \hat{J}^{(2)} x}{x^\top x}.
\label{eq:rayleigh}
\end{equation}

We show that the upper bound for HMF approximation can be easily recovered as follows.
Let us set $x = (0, 0, \ldots, 1, \ldots, 0)^\top$ in which only the $i$-th element of $x$ is equal to unity and the rest equal to zero. 
With this $x$,
\begin{equation}
x^\top \hat{J}^{(2)} x = \beta \left(\frac{k_i}{\meank}\rho \right)^{\gamma} - \mu - \Di.
\end{equation}
Therefore,
\begin{equation}
\lambda_{\max}(\hat{J}^{(2)})  \geq \max_{1 \leq i \leq N} \beta \left(\frac{k_i}{\meank}\rho \right)^{\gamma} - \mu - \Di,
\end{equation}
and we obtain the sufficient condition of $\beta$ as
\begin{equation}
\beta > (\mu + \Di) \left(\frac{k_{\max}}{\meank}\rho \right)^{-\gamma} = \overline{\beta}_{\rm HMF}.
\label{eq:sufficient_arbitrary}
\end{equation}
Notice that this sufficient condition for the endemic state is equivalent to that for HMF approximation given by Eq.~\eqref{eq:sufficient_hmf}~\cite{Saldana2008,Juher2009,Saldana2010,Juher2014}.

\subsubsection{Improved upper bound and the rich-club phenomenon}
In this section, we show an improved upper bound of $\betac$ and argue its relationship with the so-called rich-club phenomenon observed in real networks.
Our idea of improvement stems from the observation that the upper bound given by \eref{sufficient_arbitrary} is achieved via approximation of the principal eigenvector of $\hat{J}^{(2)}$ with a vector of zeros except for the element corresponding to a single node with $k_{\max}$.
We relax the criterion about node degree and approximate the principal eigenvector with a normalized vector whose element increases with $k_i$ of the corresponding nodes.
Let us set $x$ such that
\begin{align}
x_i =
\begin{cases}
0 & (k_i < k_c),\\
\sqrt{\mathstrut \dfrac{k_i}{K_c} } & (k_i \geq k_c),
\end{cases}
\end{align}
where the free parameter $k_c$ takes an integer value in $[1, k_{\max}]$ and $K_c \equiv \sum_{j: k_j \geq k_c} k_j$ is a normalization constant. 
With this choice of $x$,
\begin{align}
x^\top \hat{J}^{(2)} x &=
\sum_{i: k_i \geq k_c} \beta \left(\frac{k_i}{\meank}\rho \right)^{\gamma} \frac{k_i}{K_c} - \mu - \Di \left( 1 - \frac{2E_{c}}{K_c} \right),\\
&\geq \beta \left(\frac{k_c}{\meank}\rho \right)^{\gamma} - \mu - \Di \left( 1 - \frac{2E_{c}}{K_c} \right),
\end{align}
where $E_c$ is the total number of links between nodes with $k_i \geq k_c$.
Therefore,
\begin{equation}
\lambda_{\max} ( \hat{J}^{(2)} ) = \max_{|x|=1} \frac{x^\top \hat{J}^{(2)} x}{x^\top x} 
\geq \max_{k_c} \left[ \beta \left(\frac{k_c}{\meank}\rho \right)^{\gamma} + \Di \frac{2E_{c}}{K_c} \right] - \left( \mu + \Di \right).
\end{equation}
A sufficient condition for the endemic equilibrium is that there exists $k_c$ which satisfies
\begin{equation}
\beta \left(\frac{k_c}{\meank}\rho \right)^{\gamma} - \left( 1-\frac{2E_{c}}{K_c} \right) \Di  - \mu > 0.
\end{equation}
In terms of $\beta$, this is equivalent to
\begin{equation}
\beta > \overline{\beta} \equiv \min_{k_c} \left\{ \left[ \mu + \left( 1-  \frac{2 E_{c}}{K_{c}} \right) \Di \right] \left(\frac{k_c}{\meank}\rho \right)^{-\gamma} \right\}. 
\label{eq:sufficient_richclub}
\end{equation}
This sufficient condition always improves the previous one (given by \eref{sufficient_arbitrary})
in the following sense. Let us focus on the right hand side of \eref{sufficient_richclub} with $k_c = k_{\max}$. If there is no link between the nodes with $k_i = k_{\max}$, then $E_c = 0$ and we recover \eref{sufficient_arbitrary}. Otherwise, if $E_c > 0$, the r.h.s. of \eref{sufficient_richclub} is smaller than the r.h.s. of \eref{sufficient_arbitrary} and we obtain a smaller upper bound.
We denote the right hand side of \eref{sufficient_richclub} by $\overline{\beta}$, as an improved upper bound of endemic threshold to be compared with $\overline{\beta}_{\rm HMF}$.

The expression on the right hand side of \eref{sufficient_richclub} is closely related to so-called rich-club phenomenon of networks~\cite{Zhou2004,Colizza2006}, characterized with a high connectivity between nodes with large degree, and to the presence of a $k$-core \cite{Dorogovtsev2006}, as we discuss further below. 
Formally, the rich-club phenomenon is defined by focusing on \cite{Zhou2004,Colizza2006}
\begin{align}
\phi(k) \equiv \frac{2E_{> k}}{N_{> k} (N_{> k}-1)},
\end{align}
where $N_{> k}$ is the number of nodes with degree larger than $k$ and $E_{> k}$ is the total number of links between these nodes.
The value of $\phi(k)$ for the original network is usually compared to $\phi_{\rm rand}(k)$ of a network generated by randomly rewiring links while preserving degree of each node.
If $\phi(k) > \phi_{\rm rand}(k)$ for large values of of $k$, the network is said to exhibit the rich-club phenomenon~\cite{Colizza2006}.
The factor $2E_c / K_c$ which appears in Eq.~\eqref{eq:sufficient_richclub} is similar to $\phi(k)$ and we expect that $\overline{\beta}$ would estimate $\beta_c$ better especially when the network has the rich-club structure (see Sec.~\ref{sec:airport} for a real network example).
It should be noted that our derivation of Eq.~\eqref{eq:sufficient_richclub} results from the stability analysis of the dynamical process while $\phi(k)$ is purely based on network structure. 
The sufficient condition $\overline{\beta}$ depends on the rate of contacts inside high degree nodes, as previously, but also on the structure of the network, via the existence of links between these high-degree nodes.
The extent to which $\overline{\beta}$ improves $\overline{\beta}_{\rm HMF}$ depends on the underlying network. It is also important to stress that structure becomes more and more important as $D_I$ is increased, and the disease rapidly mixes in the population.

It is worth noting the relationship between our upper bound $\overline{\beta}$ given by \eref{sufficient_richclub} and another upper bound recently derived based on a mean-field approximation taking into account the correlation of degrees between adjacent nodes~\cite{Juher2014},
\begin{equation}
\beta > \overline{\beta}_{\rm MFC} \equiv \min_{k_i} \left\{ \left[ \mu + \left( 1-  P(k_i | k_i) \right) \Di \right] \left(\frac{k_i}{\meank}\rho \right)^{-\gamma} \right\},
\label{eq:sufficient_mfc}
\end{equation}
where the subscript MFC stands for mean-field approximation with correlated structure. 
If $P(k_c | k_c) = 2 E_c / K_c$ holds true for the $k_c$  minimizing both Eqs.~\eqref{eq:sufficient_richclub} and \eqref{eq:sufficient_mfc}, $\overline{\beta}_{\rm MFC}$ is equivalent to $\overline{\beta}$.
The accuracy of each upper bound is expected to vary depending on the structural correlations present in the network.
It should be noted that the derivation of $\overline{\beta}_{\rm MFC}$~\cite{Juher2014} assumes a mean-field approximation of the dynamical process, where nodes with the same $k_i$ value behave in the same manner (as we saw in Eqs.~\eqref{eq:rhosk_mean-field} and \eqref{eq:rhoik_mean-field} in Sec.~\ref{sec:hmf}). 
In contrast, our derivation of $\overline{\beta}$ is directly derived from Eq.(\ref{eq:rhoii}) and is free from such an approximation.
It  should also be noted that $\overline{\beta}_{\rm MFC}$ does not take into account, by definition,  structural correlations beyond those of the degrees between adjacent nodes,  and that $\overline{\beta}_{\rm MFC}$ is determined by point correlation $P(k_i | k_i)$.
The upper bound $\overline{\beta}$ is also affected by degree correlations, but it depends on the global structure of the network, as the rich-club phenomenon, because correlations are considered in a range of $k_i$.
We will compare the accuracy of $\overline{\beta}$ and $\overline{\beta}_{\rm MFC}$ in the next section for both synthetic and empirical networks.

\subsection{Numerical results}
In this section, we investigate the accuracy of the upper bounds of $\betac$ (that we obtained and reviewed in Sec.~\ref{sec:arbitrary}) by comparing them with numerical simulations of the epidemic dynamics.
For the sake of simplicity, we assume $\Ds = \Di$ and fix $(\mu, \rho)=(1, 100)$.
We focus on three parameters $(\beta, \gamma, \Di)$ and calculate $f_{\rm I}^* \equiv \sum_i \rhoii^* / n_p$, which is the fraction of infected particles at the steady state, for different parameter values.
For all numerical simulations, we set the initial condition to $(\rhosi, \rhoii) = (k_i / \meank \rho, 0)$ $(1 \leq i \leq N)$ except for the initially infected node~$j$ which is chosen at random and assigned with $\rhoij =4$ and $\rho_j = k_j / \meank \rho$. 
Even above the endemic threshold, the SIS dynamics can hit the absorbing state, \ie, the disease-free equilibrium with non-zero probability, due to stochastic fluctuations.
To circumvent this issue and compute the stationary state, we adapt the so-called quasi-stationary simulation method~\cite{Oliveira2005,Ferreira2011,Mata2013} in which, every time the system reaches the absorbing state, the configuration of the nodes' states is replaced with one chosen randomly from the history.
We also implement the simulation as it is done for migration processes~\cite{Barthelemy2010}, by not distinguishing particles and keeping only $\left( \rhosi, \rhoii \right)$ for each node.
The time evolution is simulated as a set of transitions of the Markov chain in which a state corresponds to a  configuration $\left\{ \left( \rho_{{\rm S},1}, \rho_{{\rm I},1} \right), \ldots, \left( \rho_{{\rm S},N}, \rho_{{\rm I},N} \right) \right\}$~\cite{Barthelemy2010}.
For each combination of parameters' values, we calculate $f_{\rm I}^*$ as follows. We run the epidemic process from $t=0$ to $t=T \equiv 1000/\Di$, which is $1000$ times longer than the characteristic time scale of particles' migration. In this period, $\rho_i$ converges to its stationary distribution. Then we take average of $f_{\rm I}(t) \equiv \sum_i \rhoii(t) / n_p$ over $T/2 \leq t \leq T$ and regard this average as the stationary value $f_{\rm I}^*$.

This section consists of the following four parts.
In Sec.~\ref{sec:rrg_lattice}, we consider the cases of a random regular graph with $k=4$ and a square lattice in order to confirm that regular networks exhibit the same behavior, independently of their internal organization.
In Sec.~\ref{sec:wheel}, we consider the case of the wheel graph in order to show that the sufficient condition holds true for a network whose structure differs from that of random graphs.
In Sec.~\ref{sec:bimodal}, we consider the case of random bimodal graphs with degree correlations, where $k_i$ takes either of two values and where the degree correlation is given by $P(k_i | k_j)$, in order to systematically investigate  the accuracy of the upper bounds in the presence of correlations.
Finally, in Sec.~\ref{sec:airport}, we consider the case of the US airports network, and show that the endemic threshold for a real-world rich-club network is closer to $\overline{\beta}$ than to $\overline{\beta}_{\rm HMF}$ and $\overline{\beta}_{\rm MFC}$. 
 
\subsubsection{Random regular graph and square lattice}\label{sec:rrg_lattice}
\begin{figure}
\centering
\includegraphics[width=0.45\hsize]{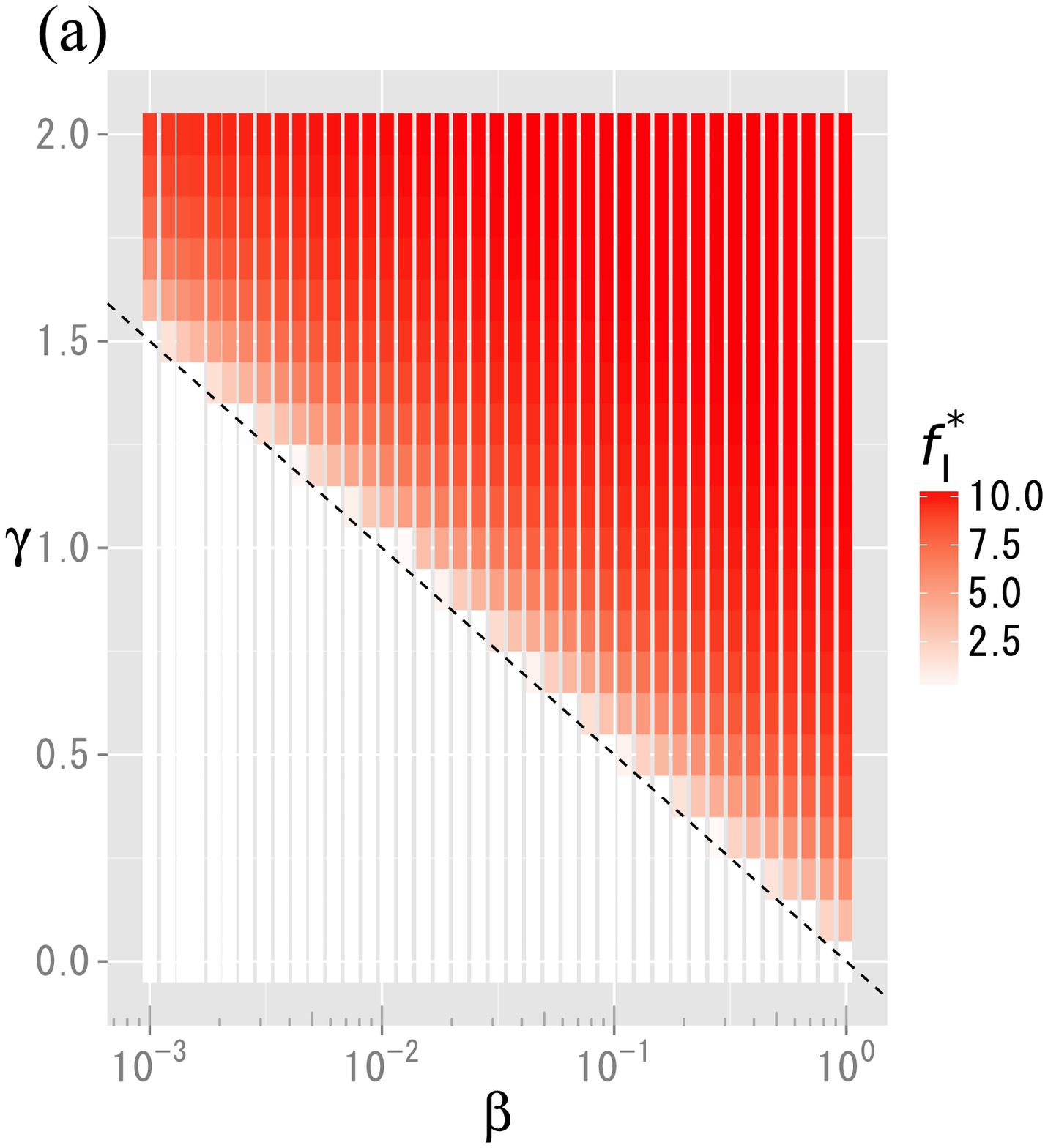}
\includegraphics[width=0.45\hsize]{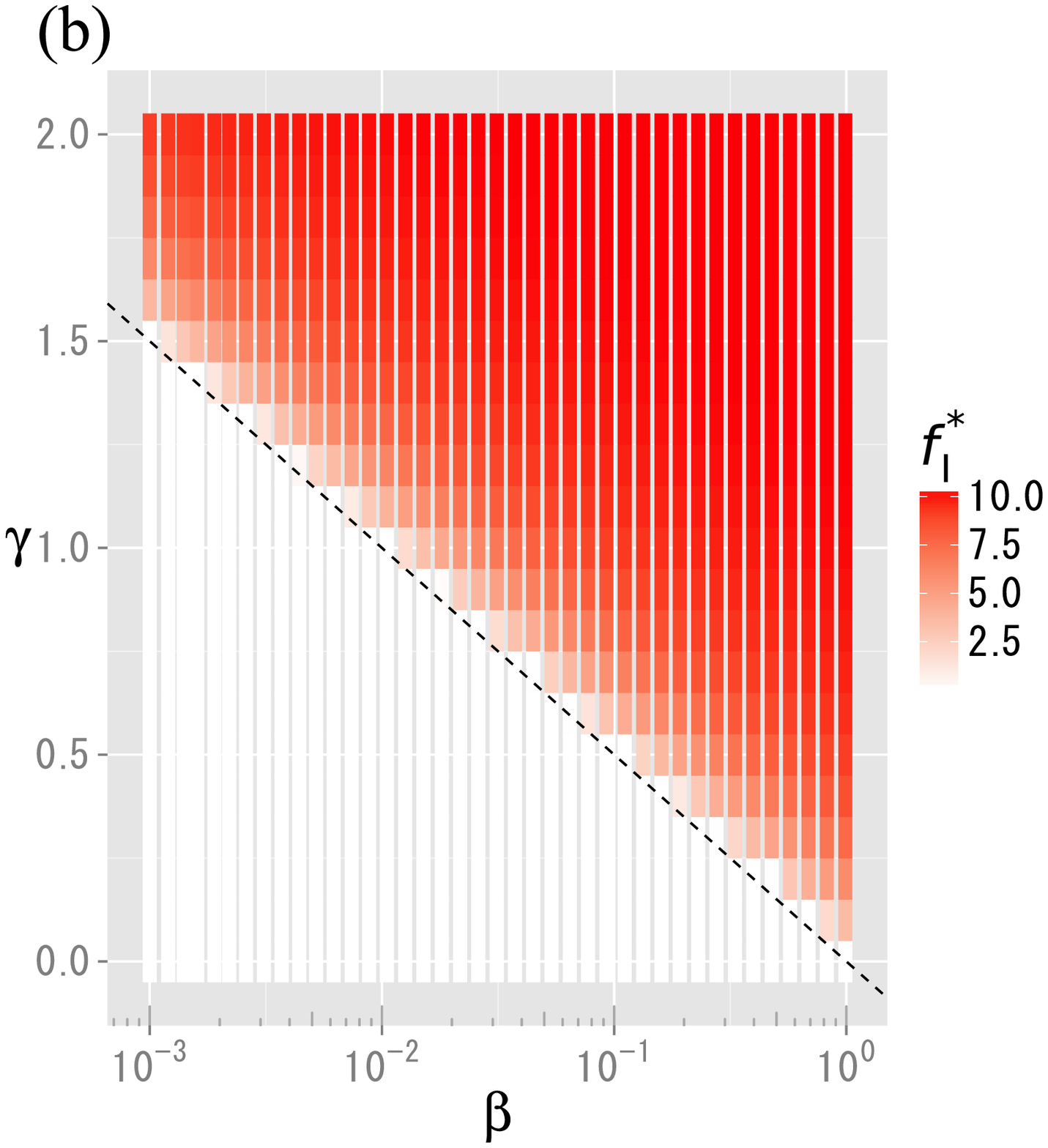}
\includegraphics[width=0.45\hsize]{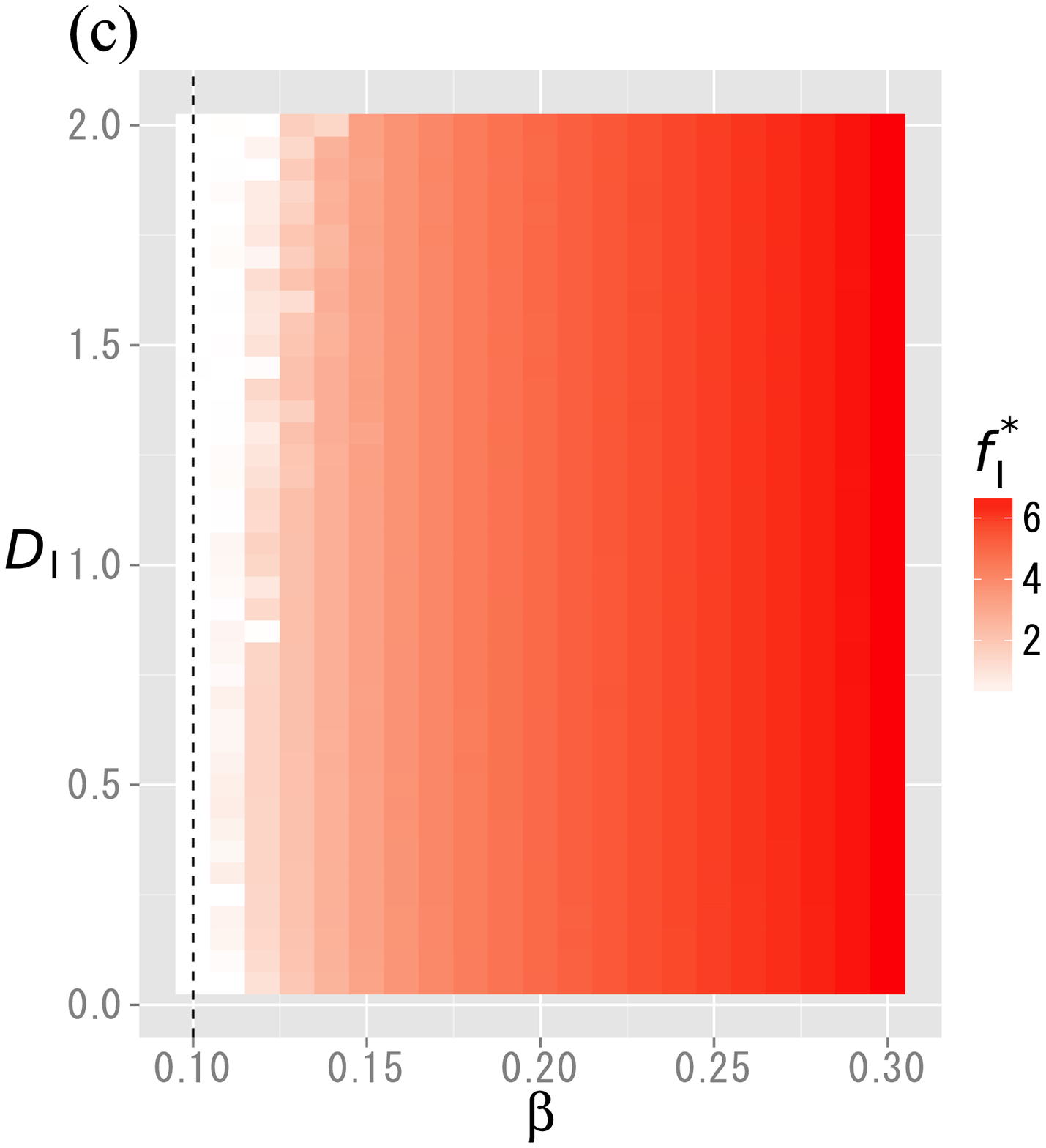}
\includegraphics[width=0.45\hsize]{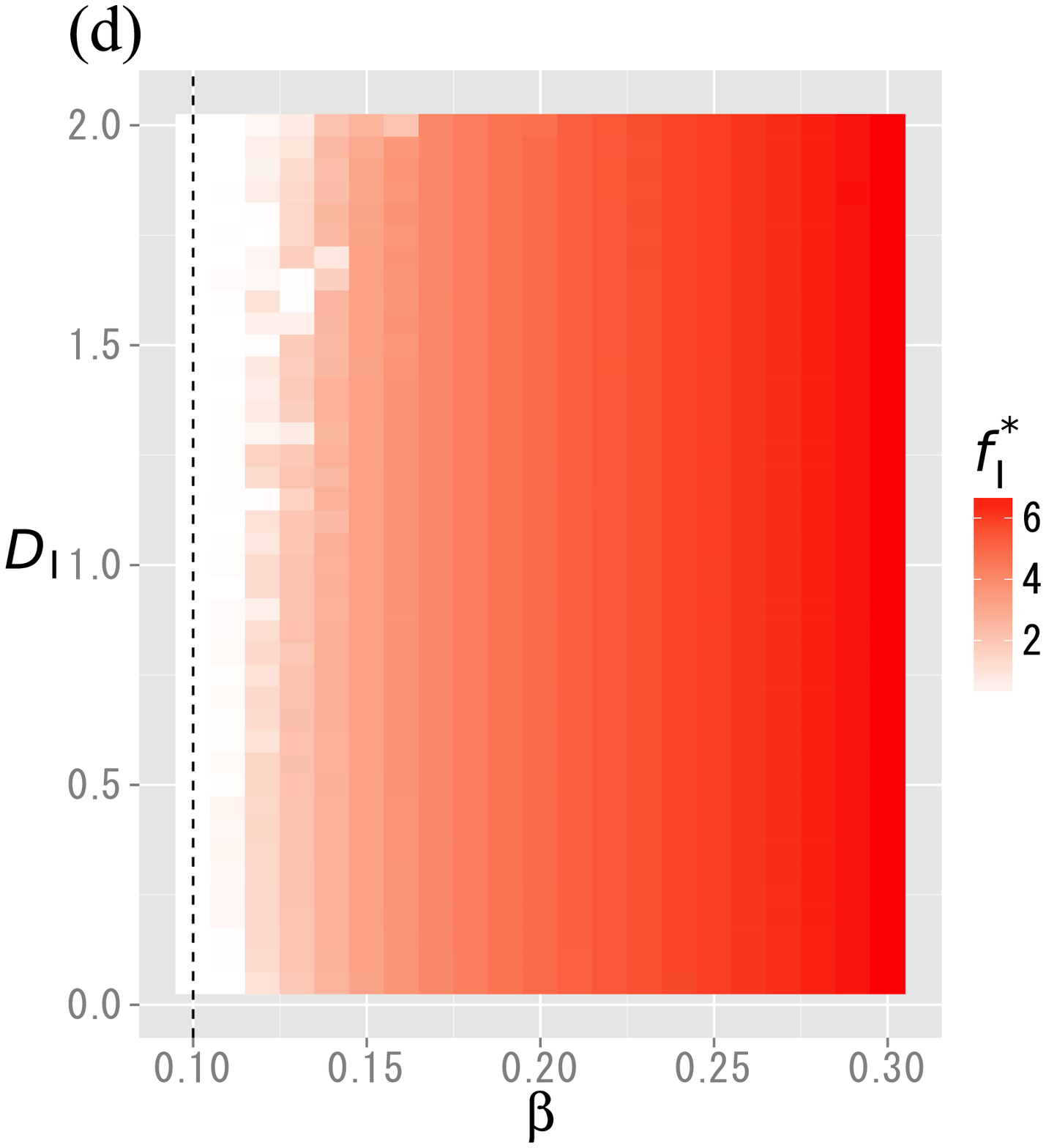}
\caption{Results of numerical simulations on (a),(c) the random regular graph with $k=4$ and on (b),(d) the square lattice, both with $N=1024$.
Fraction of infected particles at the stationary state, denoted by $\fia$, are plotted as color maps on (a),(b) $(\beta, \gamma)$-plane and on (c),(d) $(\beta, \Di)$-plane.
We set $\Di=1$ for (a),(c), and $\gamma=0.5$ for (b),(d).
The dotted lines represent the theoretical prediction of $\betac = \mu \rho^{-\gamma}$.
}
\label{fig:rrg_lattice}
\end{figure}

We generate an instance of the random regular graph with degree $k=4$ with the use of configuration model~\cite{Molloy1995,Newman2010}.
We also use the square lattice with periodic boundary condition in both directions, \ie, the lattice on the surface of a torus. 
We set $N=1024$ for both networks.
In \figref{rrg_lattice}, the fraction of infected particles at the stationary state, \ie, $\fia$, is plotted with color maps as a function of focal parameters $(\beta, \gamma, \Di)$.
Figures~\ref{fig:rrg_lattice}(a) and \ref{fig:rrg_lattice}(b) show the results on the $(\beta,\gamma)$-plane.
Although we observe deviation in the $\fia$ values due to the fluctuation of stochastic simulations, the theoretical prediction of $\betac =\mu \rho^{-\gamma}$ (drawn as the dotted line) fits well with the simulation results.
Figures~\ref{fig:rrg_lattice}(c) and \ref{fig:rrg_lattice}(d) show the results on the $(\beta,\Di)$-plane.
The simulation results indicate that the onset of the endemic state does not depend on $\Di$, which is consistent with the theoretical prediction $\betac =\mu \rho^{-\gamma}$.
In those figures, the regular random graph and square lattice are quantitatively indistinguishable, despite the fact that the former is appropriate for a mean-field approximation and the latter not.

\subsubsection{Wheel graph}\label{sec:wheel}
\begin{figure}
\centering
\includegraphics[width=0.45\hsize]{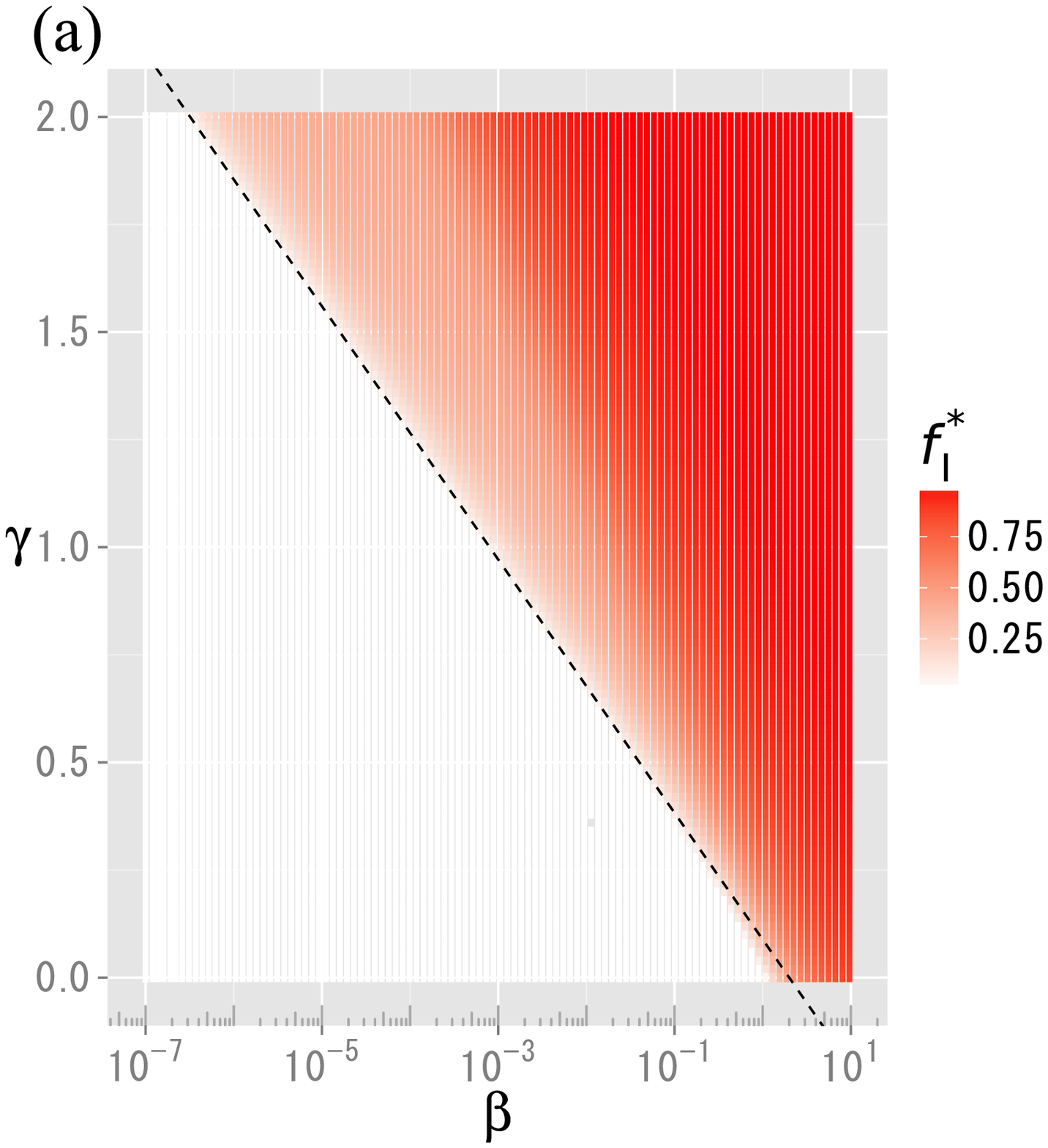}
\includegraphics[width=0.45\hsize]{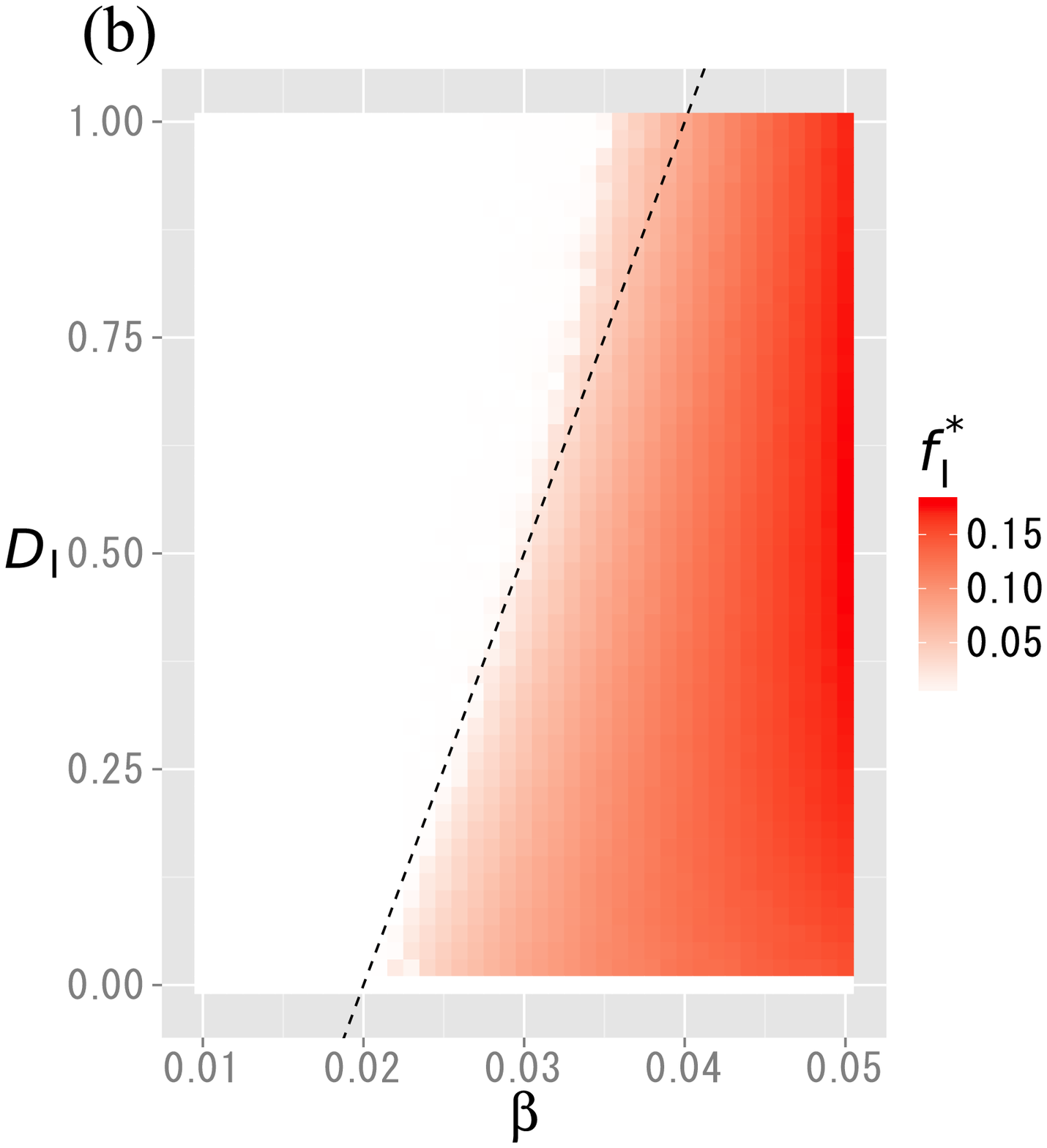}
\caption{
Results of numerical simulations on the wheel graph with $N=100$.
Fraction of infected particles at the stationary state, denoted by $\fia$, are plotted as color maps on (a) $(\beta, \gamma)$-plane and on (b) $(\beta, \Di)$-plane.
We set $\Di = 1$ for (a) and $\gamma = 0.5$ for (b). 
Dotted lines represent the upper bound of $\betac$ given by $\overline{\beta} =  (\mu + \Di)((N-1) / {\meank}\rho)^{-\gamma}$.
}
\label{fig:beta_gamma_D}
\end{figure}

The wheel graph with $N$ nodes consists of a cycle with $N-1$ nodes and a single hub connected to the rest of the nodes.
Nodes on the cycle have degree~$3$ and the node at the center has degree~$N-1$.
The average degree of the $N$-wheel graph is equal to $\meank = 4(N-1)/N$.
For the wheel graph, $\overline{\beta} = \overline{\beta}_{\rm HMF} = \overline{\beta}_{\rm MFC}$ holds true and are equal to $(\mu + \Di) \left( (N-1) / \meank \rho \right)^{-\gamma}$ if $N \gg 1$.
In Fig.~\ref{fig:beta_gamma_D}(a), $\fia$ is plotted on the $(\beta, \gamma)$-plane with $\Di=1$.
The upper bound is shown as the dashed line
and is very accurate except for $0 \leq \gamma \lesssim 0.2$.
In Fig.~\ref{fig:beta_gamma_D}(b), $\fia$ is plotted on the $(\beta, \Di)$-plane with $\gamma=0.5$.
The upper bound is shown as the dashed line again but is not tight when $0.6 \lesssim D < 1$.
As a summary, the sufficient conditions given by Eqs.~\eqref{eq:sufficient_arbitrary}, \eqref{eq:sufficient_richclub}, and \eqref{eq:sufficient_mfc} provide accurate description of the epidemic threshold
even for a network where a mean-field assumption is not expected to hold.

\subsubsection{Random bimodal graphs with degree correlation}\label{sec:bimodal}

In the previous sections, we have focused on random graphs with no correlation and on more regular graphs.
However, real-world networks are neither purely random nor deterministic but often exhibit structural correlations.
A typical example of structural correlation is the correlation of the degrees between adjacent node pairs, so-called degree assortativity \cite{Newman2002,Newman2003}.
If two adjacent nodes tend to have similar degree values, the network is called assortative.
If the nodes with a large degree tend to be connected to the nodes with a small degree, the network is instead called disassortative. Empirical observations suggest that similar types of networks are characterized by similar levels of assortativity~\cite{Newman2010}.
The degree assortativity is quantified by the degree assortativity coefficient $r$ which is defined as the Pearson correlation coefficient of the degrees of adjacent nodes \cite{Newman2002,Newman2003} and has values in the range $-1 \leq r \leq 1$.
A positive (negative) $r$ value represents an assortative (disassortative) network.
Determining the influence of degree assortativity on the accuracy of the upper bounds $\overline{\beta}_{\rm HMF}$, $\overline{\beta}$, and $\overline{\beta}_{\rm MFC}$ is the main purpose of this section.

In order to tune the assortativity present in the networks,  we generate instances of random bimodal graph with degree correlations as follows.
In the random bimodal graph, a node has its degree equal to either of two values denoted by $k_1$ and $k_2$ $(k_1 > k_2)$, with probability $a$ and $1-a$, respectively.
The degree assortativity coefficient is determined by the parameter $P(k_1 | k_2)$ which is the probability of a random link to have a node with degree $k_1$ as an end under the condition that the other end has $k_2$~\cite{Hasegawa2013}, and equal to $r = 1 - (a k_1 + (1-a) k_2) P(k_1 | k_2) / a k_1$ (see Appendix A for the derivation).
We also calculate the expecting values of $\overline{\beta}$ and $\overline{\beta}_{\rm MFC}$ as (see Appendix B for the derivations),
\begin{align}
\overline{\beta} &= {\rm min}  \left[ \left( \mu + (1-r) \frac{(1-a)k_2}{\langle k \rangle} \Di \right) \left( \frac{k_1}{\langle k \rangle} \rho \right)^{-\gamma}, \mu \left( \frac{k_2}{\langle k \rangle} \rho \right)^{-\gamma} \right],\\
\overline{\beta}_{\rm MFC} &= {\rm min}  \left[ \left( \mu + (1-r) \frac{(1-a)k_2}{\langle k \rangle} D_{\rm I} \right) \left( \frac{k_1}{\langle k \rangle} \rho \right)^{-\gamma}, \left( \mu + (1-r) \frac{a k_1}{\langle k \rangle} \Di \right) \left( \frac{k_2}{\langle k \rangle} \rho \right)^{-\gamma} \right].
\end{align}
Because $(1-r) a k_1 \Di / \langle k \rangle \geq 0$, $\overline{\beta} \leq \overline{\beta}_{\rm MFC}$ holds true regardless of parameter settings.
In other words, $\overline{\beta}$ is eqivalent to or better than $\overline{\beta}_{\rm MFC}$ as an upper bound of $\beta_c$ for the random bimodal graph.

\begin{figure}
\centering
\includegraphics[width=0.325\hsize]{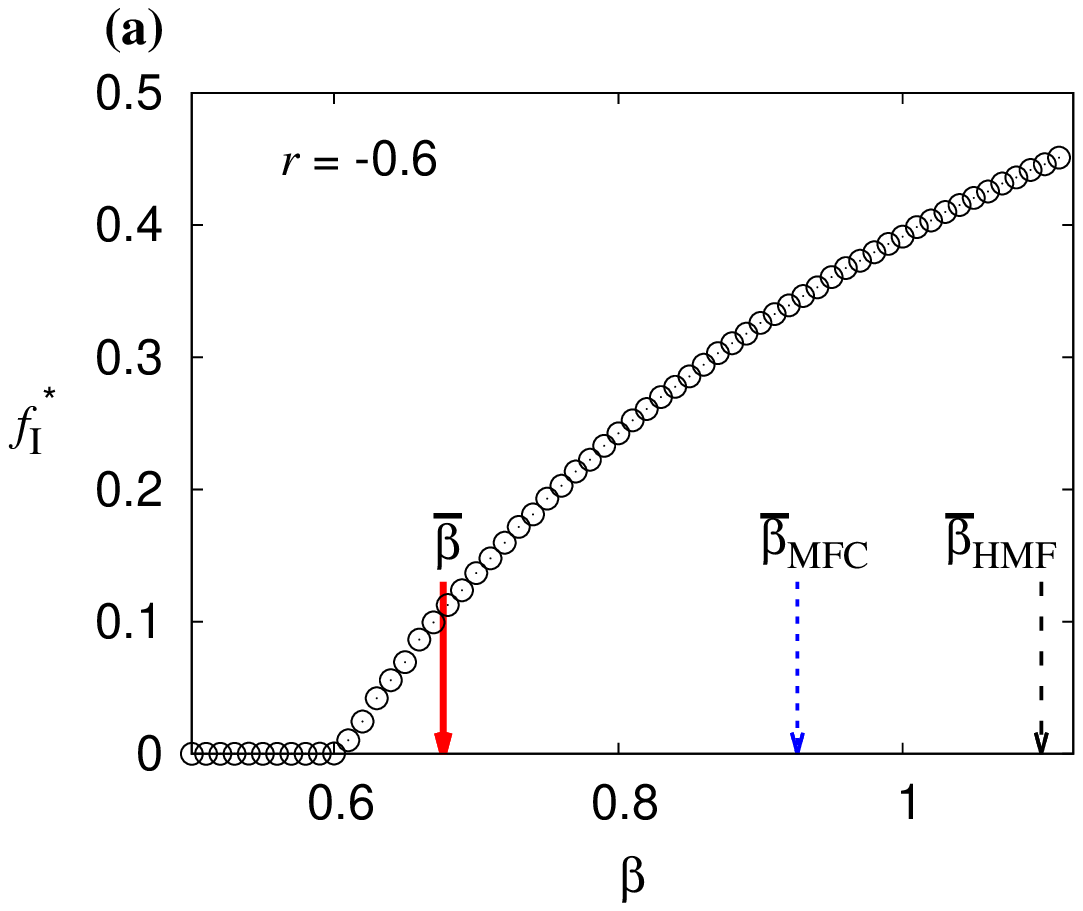}
\includegraphics[width=0.325\hsize]{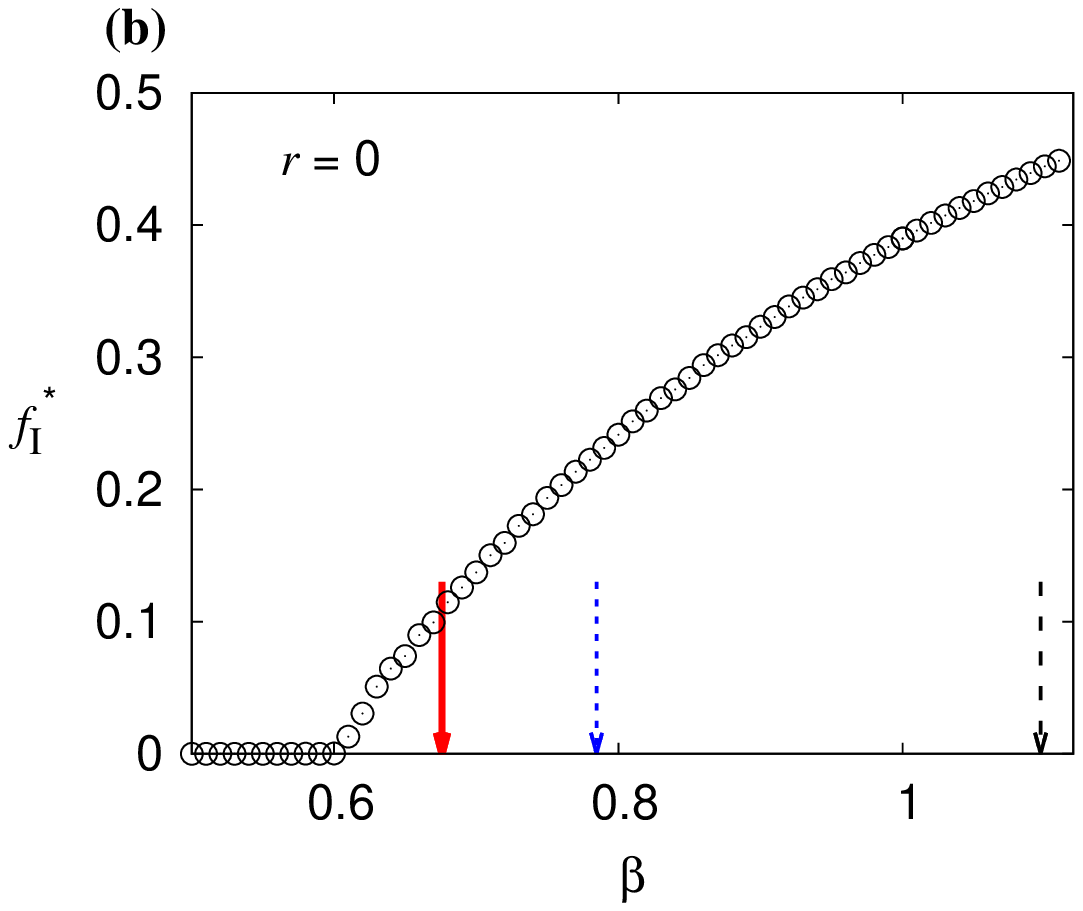}
\includegraphics[width=0.325\hsize]{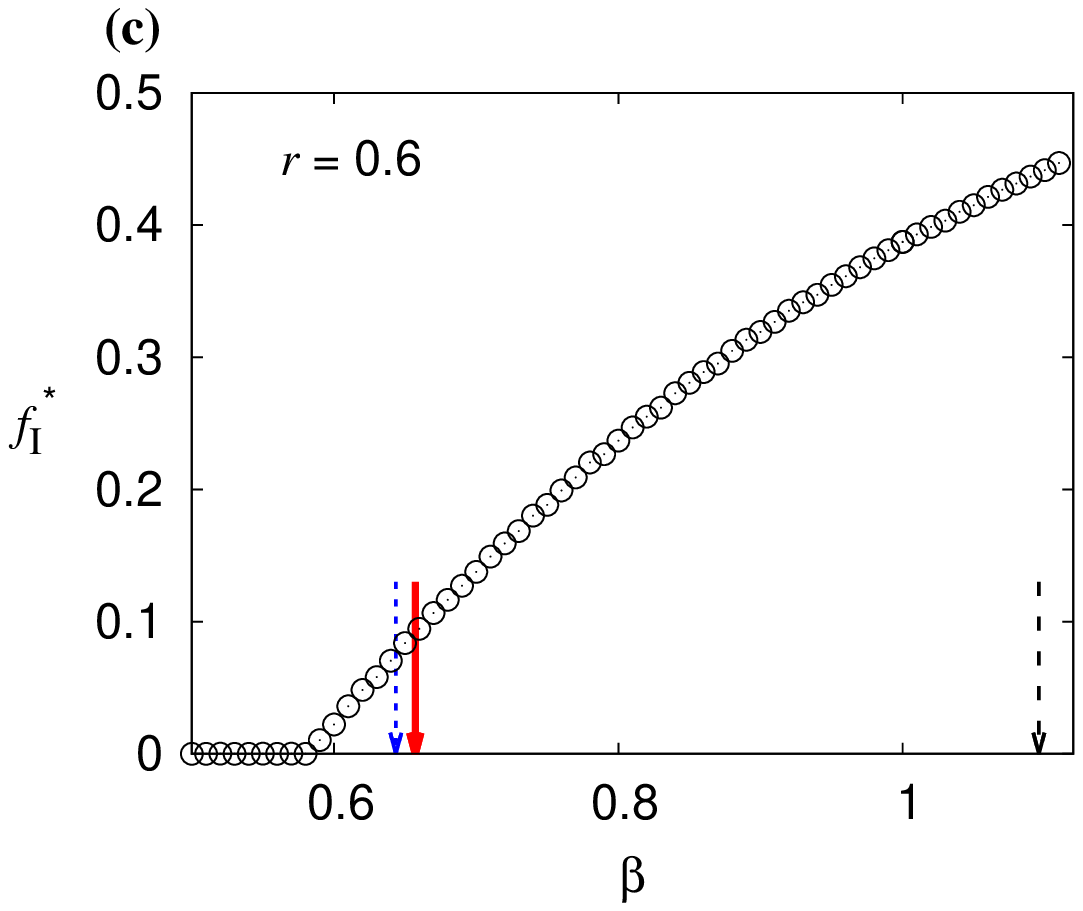}\\
\includegraphics[width=0.325\hsize]{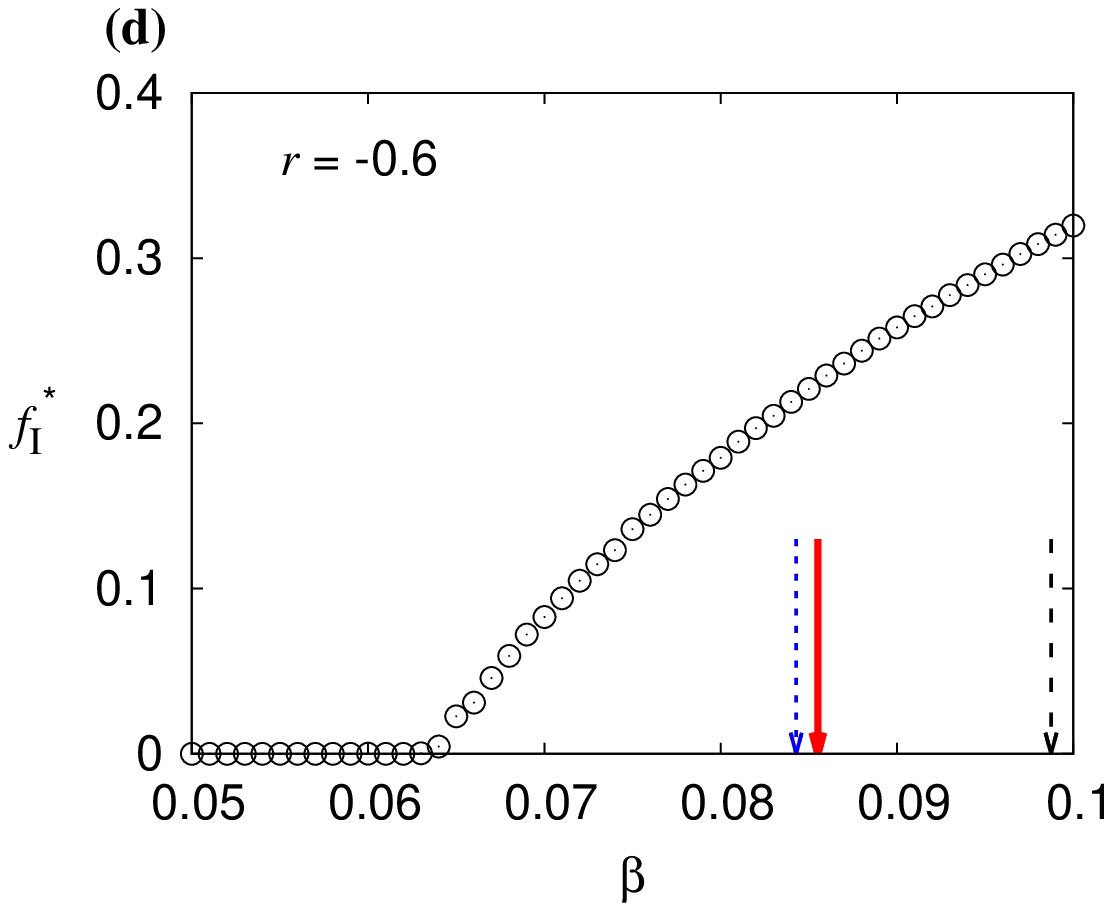}
\includegraphics[width=0.325\hsize]{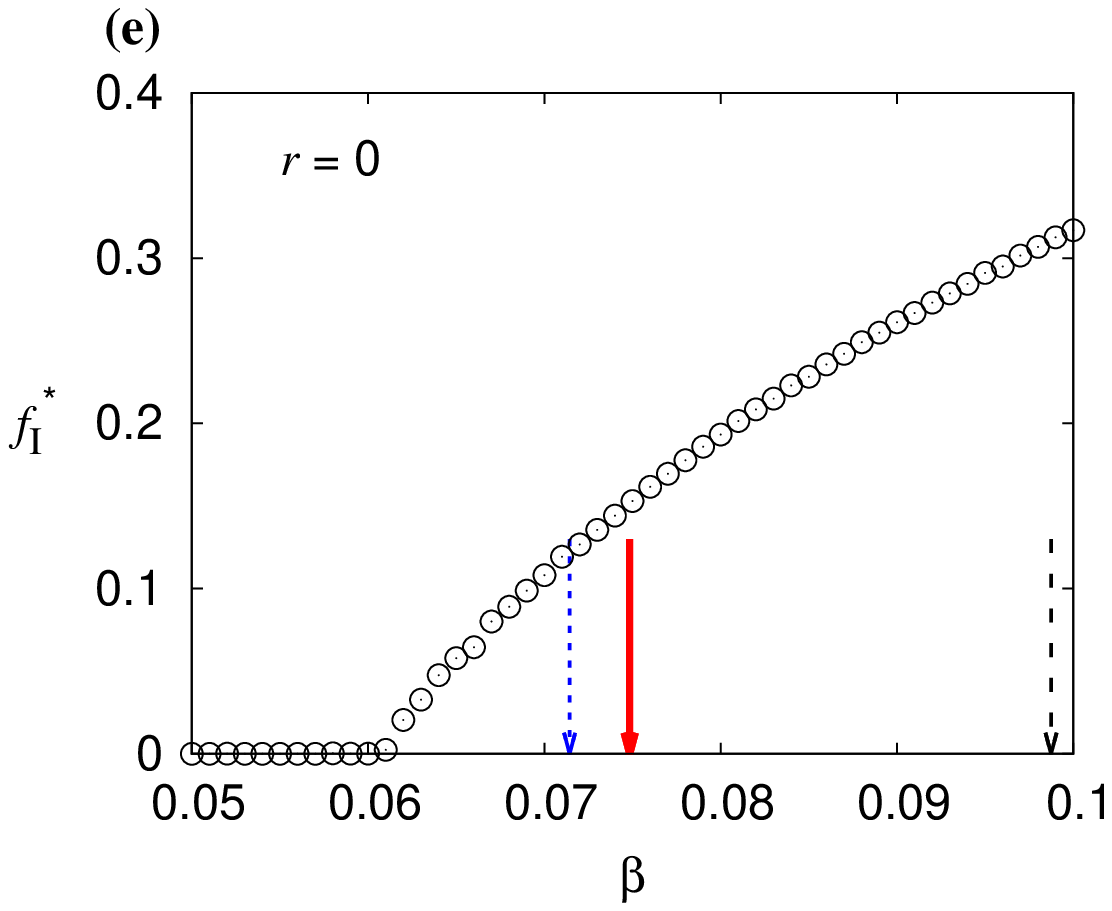}
\includegraphics[width=0.325\hsize]{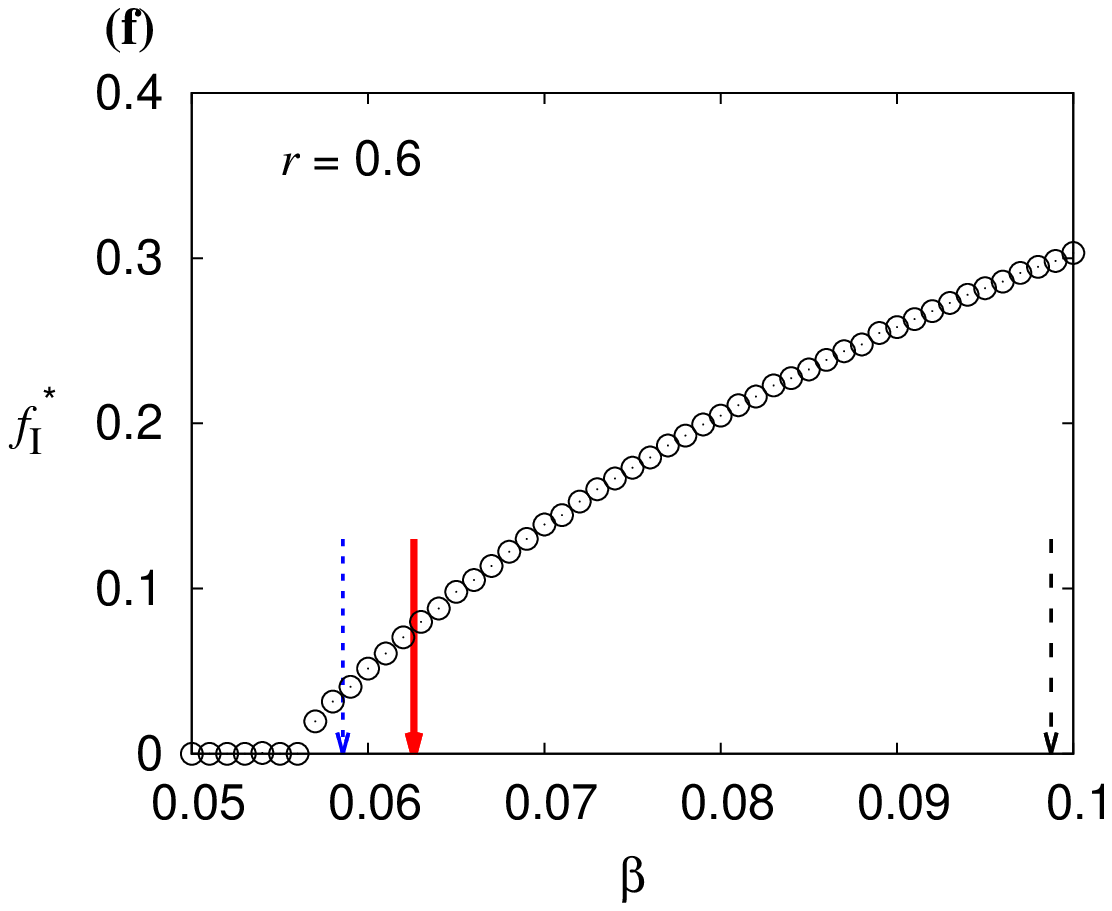}
\caption{
Results of numerical simulations on the random bimodal graph with degree correlation with $(N, k_1, k_2, a) = (1000, \ 16, \ 2, \ 1/7)$.
Fraction of infected particle at the stationary state, denoted by $\fia$, are plotted as a function of $\beta$ for (a),(b),(c) $\gamma=0.1$ and (d),(e),(f) $\gamma = 0.5$.
We set the degree assortativity coefficient $r$ to (a),(d) $-0.6$, (b),(e) $0$, and (c),(f) $0.6$.
The vertical arrows indicate the three upper bounds: The solid ones are $\overline{\beta}$, the dotted ones $\overline{\beta}_{\rm MFC}$, and the dashed ones $\overline{\beta}_{\rm HMF}$.}
\label{fig:bimodal}
\end{figure}

In Fig.~\ref{fig:bimodal}, $\fia$ is plotted as a function of $\beta$ for random bimodal graphs with disassortative, neutral, and assortative degree correlations.
We consider two values of $\gamma$, $\gamma = 0.1$ and $0.5$.
For $\gamma =0.1$, that is when the dependency of infection rate on population size is weak, $\overline{\beta}$ provides a more accurate estimation of the endemic threshold than $\overline{\beta}_{\rm MFC}$ for the disassortative and neutral cases (Figs.~\ref{fig:bimodal}(a) and \ref{fig:bimodal}(b)). For the assortative case, $\overline{\beta}$ is less accurate than, but comparable with $\overline{\beta}_{\rm MFC}$ (Fig.~\ref{fig:bimodal}(c)).
In all three cases, $\overline{\beta}_{\rm HMF}$ gives the poorest performance.
When $\gamma =0.5$ and the dependency of infection rate on population size is strong, $\overline{\beta}$ gives less accurate estimation than $\overline{\beta}_{\rm MFC}$ for all the three cases (Figs.~\ref{fig:bimodal}(d), \ref{fig:bimodal}(e), and \ref{fig:bimodal}(b)) and the difference between the two upper bounds increases with $r$. 
The accuracy of the two upper bounds increases with $r$, while $\overline{\beta}_{\rm HMF}$ remains the same regardless of the value of $r$.
The increase in accuracy of $\overline{\beta}_{\rm MFC}$ with $r$ is expected because $P(k_1|k_1)$ and $P(k_2 | k_2)$ become larger when $r$ increases.

The theoretical results predict that $\overline{\beta} < \overline{\beta}_{\rm MFC}$ for $\gamma = 0.1$ and $r \in \left\{-0.6, 0 \right\}$ and that $\overline{\beta} = \overline{\beta}_{\rm MFC}$ for $(\gamma, r) = (0.1, 0.6)$ and $\gamma = 0.5$ (see Fig.~B1 in Appendix B).
The small discrepancy between the theoretical predictions and the numerical results shown in Fig.~\ref{fig:bimodal} is probably due to the assumption behind the theory, as we assumed that the network is a typical instance of the random bimodal graph with a sufficiently large $N$. In contrast, in the numerical simulations, we calculate $\overline{\beta}$ and $\overline{\beta}_{\rm MFC}$ for a single realization with $N=1000$.
Moreover, fluctuations in the node degree arise due to formation of self-loops and multiple links between a pair of nodes when generating the graph.
Because $\overline{\beta}$ uses the exact value of degrees as the threshold, it is affected by such fluctuations.
This observation suggests that $\overline{\beta}$ theoretically provides a robust estimate in a broad variety of degree correlations, while its accuracy may be sensitive to the fluctuations due to randomness in network generation.

\subsubsection{US airports network}\label{sec:airport}
In  this section, we show  simulation results on the US airports network~\cite{Opsahl2011,Kunegis2013}, a well-known example of rich-club network~\cite{Colizza2006}.
The system is a directed network in which nodes correspond to airports in the US and directed links between nodes represent directed flights between two airports in 2010~\cite{Opsahl2011}.
We discard the direction of all the links and regard it as an undirected network.
We focus on its largest connected component with $N=1572$ nodes and $E=28235 $ links.
The degree assortativity coefficient $r$ of the focal network is equal to $-0.1134$. The rich-club structure fosters the assortative mixing in the network, but it does not always lead to a positive $r$ value of the whole network because the rich-club phenomenon pays attention only to the nodes with large $k$ values. In this sense, the relationship between the degree correlation and rich-club phenomenon is not trivial.
With numerical simulations, we examine to what extent $\overline{\beta}$ improves $\overline{\beta}_{\rm HMF}$ and $\overline{\beta}_{\rm MFC}$ in this empirical network relevant to real-world epidemics.

\begin{figure}
\centering
\includegraphics[width=0.45\hsize]{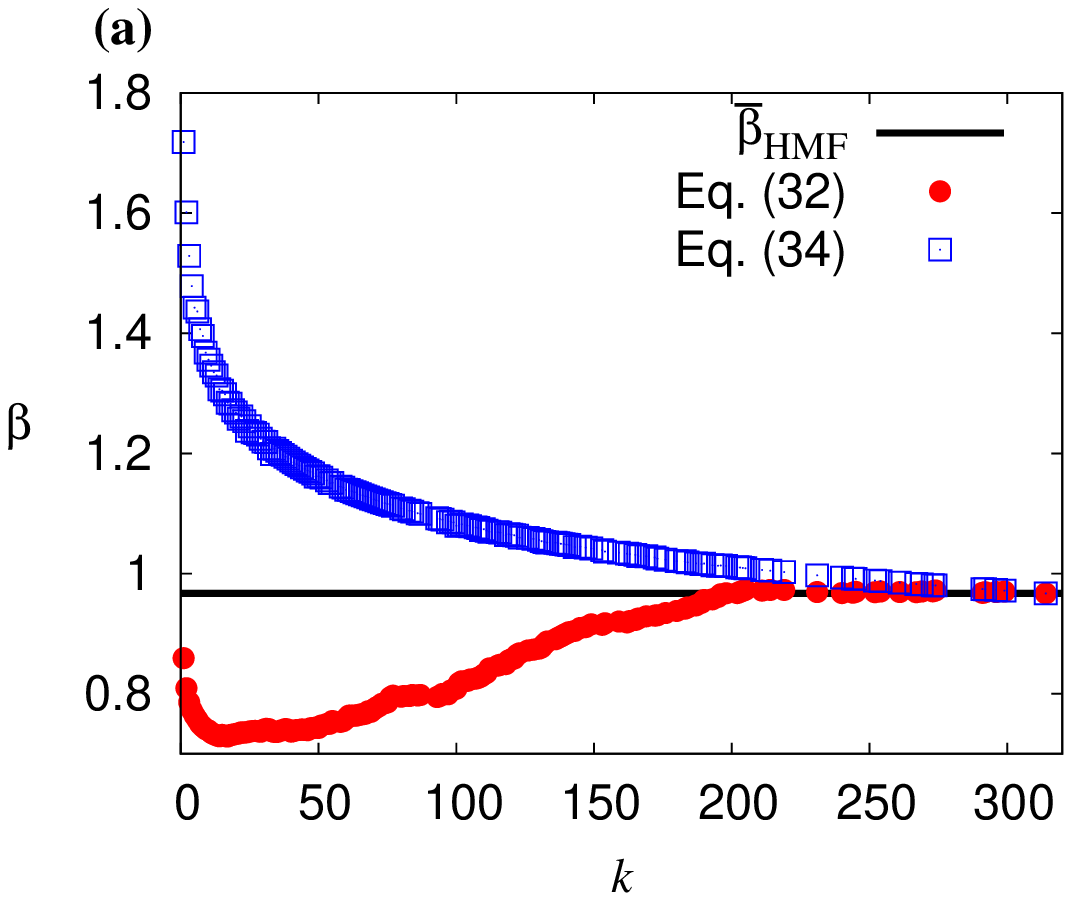}
\includegraphics[width=0.45\hsize]{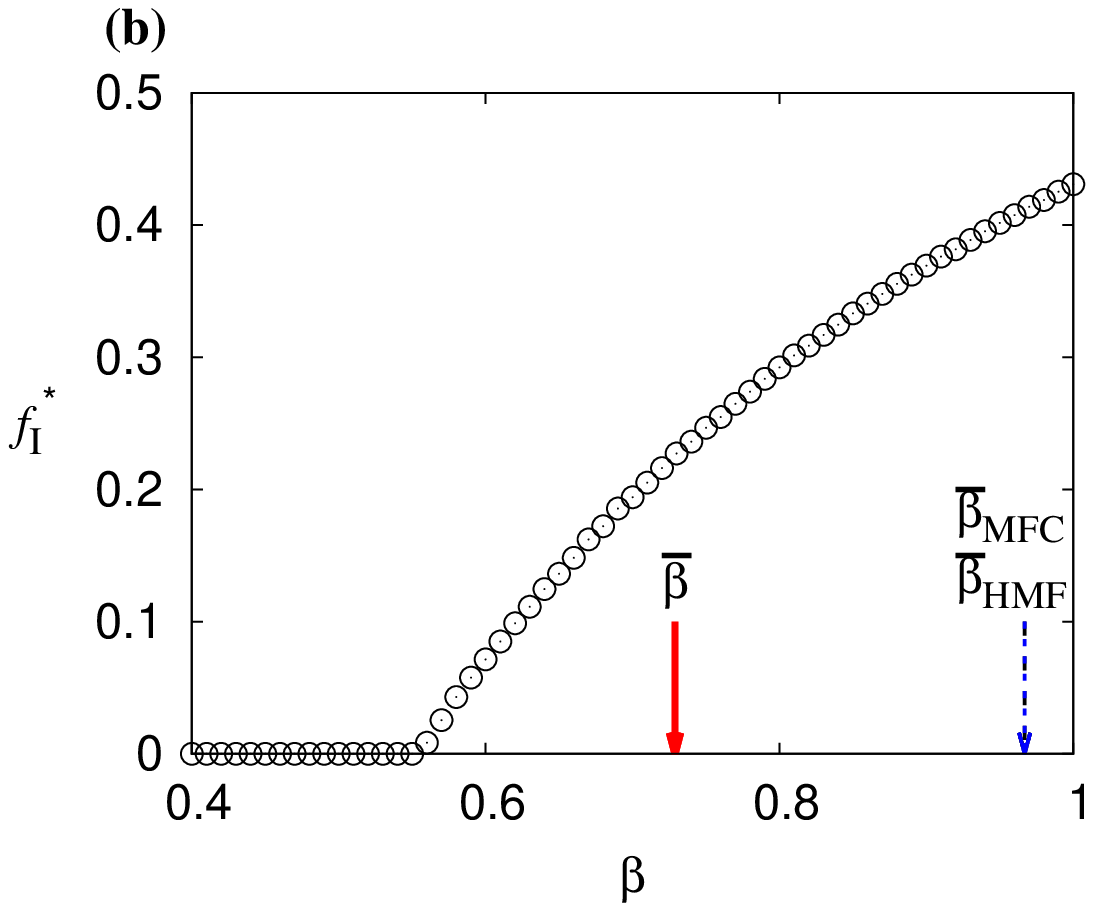}
\includegraphics[width=0.45\hsize]{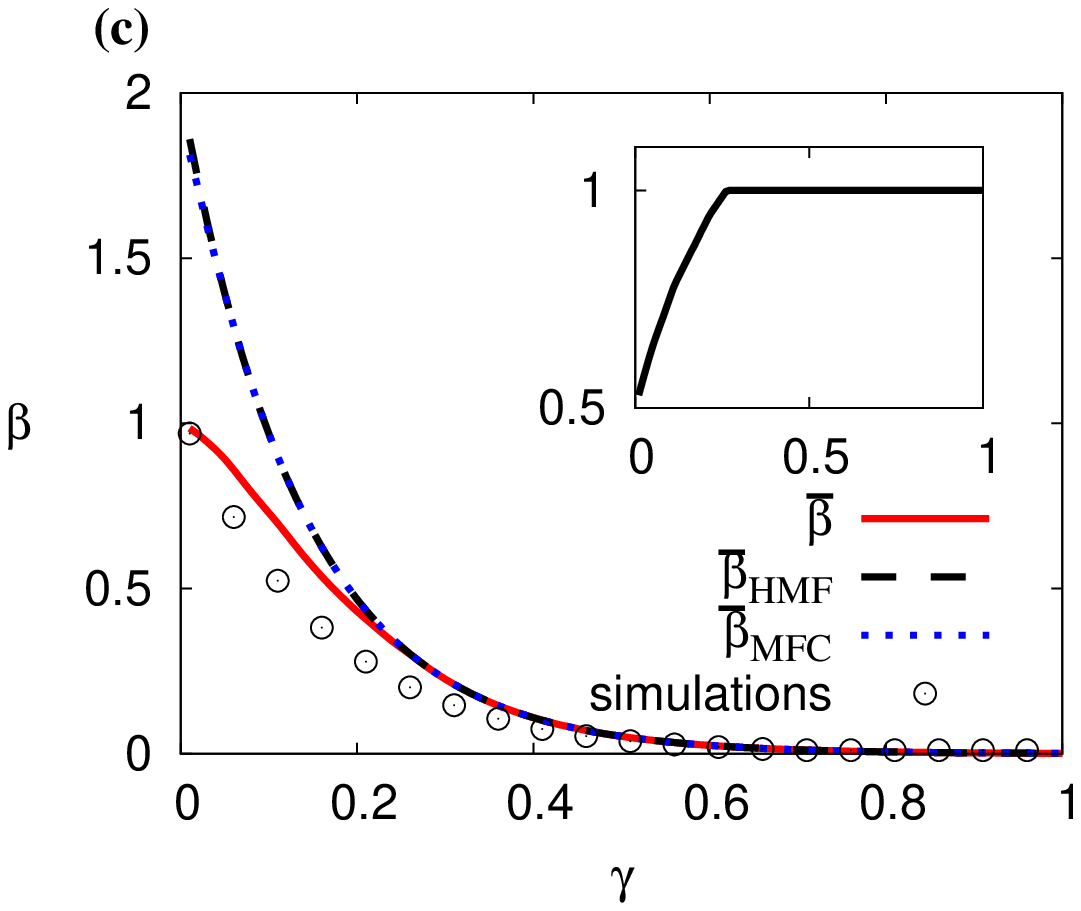}
\caption{Results of numerical simulations on the US airports network.
(a) Factor on the right hand side of \eref{sufficient_richclub} which is given by $\left( \mu + \left( 1-  2 E_{c} / K_{c} \right) \Di \right) \left(k_c \rho / \meank \right)^{-\gamma}$ as a function of $k_c$ (filled circles), and the factor on the right hand side of \eref{sufficient_mfc} which is given by $\left( \mu + \left( 1-  P(k_i|k_i) \right) \Di \right) \left(k_i \rho / \meank \right)^{-\gamma}$ as a function of $k_i$ (open squares). The solid horizontal line indicates $\overline{\beta}_{\rm HMF} = (\mu + \Di)(k_{\max} \rho / \meank)^{-\gamma}$.
(b) Fraction of infected particles at the stationary state, denoted by $\fia$, as a function of infection rate $\beta$ (open circles).
The solid, dashed, and dotted arrows point three upper bounds $\overline{\beta}$, $\overline{\beta}_{\rm HMF}$, and $\overline{\beta}_{\rm MFC}$, respectively.
We set $\Di=1$ and $\gamma=0.1$ for (a),(b).
(c) Dependency of the endemic threshold on $\gamma$.
The open circles indicate $\beta$ above which $f_{\rm I}^* > 1/\rho$ holds true for the simulations.
Inset: the ratio of $\overline{\beta}$ to $\overline{\beta}_{\rm HMF}$ shown in the main panel.
}
\label{fig:usair}
\end{figure}

In Fig.~\ref{fig:usair}(a), we first evaluate the dependence of the r.h.s. of \eqref{eq:sufficient_richclub} and \eqref{eq:sufficient_mfc} on $k$ for $\gamma = 0.1$.
The minimum value of each curve corresponds to $\overline{\beta}$ and $\overline{\beta}_{\rm MFC}$, respectively.
Because $\overline{\beta}_{\rm HMF}$ is independent of $k$, it is plotted as a horizontal line. 
While the r.h.s. of \eref{sufficient_richclub} follows a U-shape curve and takes its minimum at $k=17$, the r.h.s. of \eref{sufficient_mfc} monotonically decreases with $k$ and takes its minimum at $k=k_{\max}=314$, and its minimum value is equal to $\overline{\beta}_{\rm HMF}$.
Therefore, one finds $\overline{\beta}_{\rm MFC} = \overline{\beta}_{\rm HMF}$ for the US airports network when $\gamma = 0.1$.

In \figref{usair}(b), $\fia$ is plotted as a function of infection rate $\beta$ with $\gamma = 0.1$.
The results of numerical simulations indicate $\betac \sim 0.56$.
The vertical arrows point the three upper bounds $\overline{\beta}$, $\overline{\beta}_{\rm HMF}$, and $\overline{\beta}_{\rm MFC}$, with $\overline{\beta}_{\rm HMF}$ and $\overline{\beta}_{\rm MFC}$ overlapping, as we saw in \figref{usair}(a).
It is clear that  $\overline{\beta}_{\rm HMF}$  (equal to $0.967$) is far from the  value $\betac$ measured in the simulations, whereas $\overline{\beta}$  (equal to $0.729$) drastically improves the estimation.
These results support our main claim that  rich-club networks have a smaller upper bound for $\betac$ than non rich-club networks with the same degree sequence.

The difference between the upper bounds is then studied for values of $\gamma$  in the range $(0,1]$, as an exact expression of  $\betac$ can be derived for $\gamma=0$, and that the three upper bounds are invalid for this value (see Sec.~\ref{sec:general_case}).
In \figref{usair}(c), we plot the  values of $\betac$ estimated from the simulations and the three upper bounds are plotted as a function of $\gamma$.
We observe that $\overline{\beta}_{\rm MFC}$ is very close to $\overline{\beta}_{\rm HMF}$ over the whole the range of $\gamma$ for this network.
For $0 < \gamma \lesssim 0.2$, $\overline{\beta}$ is more accurate than $\overline{\beta}_{\rm HMF}$ and $\overline{\beta}_{\rm MFC}$ and closer to the simulation results, while the three upper bounds are equivalent for large values of $\gamma$ (see the inset of  \figref{usair}(c)).
Interestingly, real-world data tend to present small values of $\gamma$~\cite{Smith2009,Schlapfer2014}, in the regime where $\overline{\beta}$
improves most the accuracy of the estimate.
  
\section{Discussion}
In the present study, we have investigated the SIS epidemic dynamics on arbitrary metapopulation networks where the infection rate at each node depends on its population size.
We have proved that the upper bound of endemic threshold $\betac$, previously obtained on the basis of an heterogeneous mean-field approximation, holds true even for arbitrary networks, regardless of this approximation.
For an arbitrary network, the upper bound is achieved by assuming that a single node with the largest degree determines the onset of the endemic state.
In addition, we have derived an improved upper bound for the endemic threshold, which decreases when the network possesses more links between large degree nodes.
This structural property is directly related to the so-called rich-club phenomenon, observed in a variety of real-world networks.
We have verified the resulting upper bounds with numerical simulations on network models, and on an empirical network,  the US airports network, exhibiting the rich-club phenomenon and relevant to epidemic spreading.

Our finding that rich-club networks favor epidemic spreading is reminiscent of recent results showing connections between  endemic threshold and $k$-core structure for SIS dynamics on quenched contact networks (\ie, networks between individuals)~\cite{Castellano2012}.
A $k$-core in a network is defined by the subgraph remaining after repeatedly removing nodes with degree smaller than $k$ while recalculating nodes' degree on every node removal~\cite{Dorogovtsev2006}.
By definition, a $k$-core with a large $k$ represents a highly-connected subgroup of nodes with large degrees.
It was shown with  numerical simulations~\cite{Castellano2012} that nodes in the $k$-core sustain infection, a mechanism related to the finding that these nodes are influential initial spreaders for susceptible-infected-recovered (SIR) epidemic dynamics~\cite{Anderson1991} on contact networks~\cite{Kitsak2010}.
It should be noted that, prior to these physical studies, the impact of coreness in epidemic process was established in public health studies~(\eg, \cite{Laumann1999,Lowndes2002,Jolly2002}).
Although we considered metapopulation networks instead of contact networks, our finding suggests that a similar mechanism might occur in our case, as a highly-connected subgroup of nodes with large degrees sustains infection and favors onset of an endemic state.

This work opens interesting research questions related to epidemic dynamics on metapopulations networks.
First, the upper bound $\overline{\beta}$ is essentially obtained by estimating the principal eigenvector of the Jacobian matrix of the disease-free equilibrium, by using local network structure only (namely, node degree $k_i$).
As we have shown in the US airports network, this upper bound improves results obtained in the mean-field approach,
but there is still room for improvement in order to predict the actual $\betac$ measured in numerical simulations.
Further improvement could be found by taking into account higher-order structural properties, such as community structure \cite{Fortunato2010}.
Second, in this paper, we have focused on the endemic threshold solely, but other quantities describe other properties of the dynamical process, often crucial from a theoretical point of view but also in practical public health. An important example is the fraction of infected particles at the stationary state, which requires nonlinear analytical approaches because the nonlinear terms associated with $\rhosi\rhoii$ are not negligible and the linearization of the system fails in this regime.

\section*{Acknowledgments}
The authors thank to Naoki Masuda and Petter Holme for valuable discussions.
This work was partly supported by Bilateral Joint Research Projects between JSPS, Japan, and F.R.S.--FNRS, Belgium.
R.L. acknowledges support from FNRS, EU-FP7 project `Optimizr', IAP DYSCO and ARC `Mining and Optimization of Big Data Models'. 
The data of the US airports network is downloaded from the Koblenz Network Collection, \url{http://konect.uni-koblenz.de/}.

\section*{Appendix A: Degree assortativity coefficient $r$ for the random bimodal graph}
\setcounter{equation}{0}
\renewcommand{\theequation}{A\arabic{equation}}
From Eq.~(51) in Ref.~\cite{Serrano2007}, we can rewrite the degree assortativity coefficient $r$ as
\begin{align}
r = 1 - \frac{\langle k \rangle \langle k^3 \rangle - \langle k \rangle \sum_k k^2 \overline{k}_{\rm nn}(k) P(k)}{\langle k \rangle \langle k^3 \rangle - \langle k^2 \rangle^2},
\label{eq:r}
\end{align}
where $\overline{k}_{\rm nn}(k)$ is the average degree of the nearest neighbor nodes associated with a node with degree $k$,
\begin{align}
\overline{k}_{\rm nn}(k) = \sum_{k^\prime} k^\prime P(k^\prime | k).
\end{align}

We suppose that a node in the random bimodal graph has degree $k_1$ and $k_2$ with probability $p(k_1) = a$ and $p(k_2) = 1-a$, respectively ($k_1 \geq k_2$).  
The degree correlation is governed by the conditional probability $p(k_1 | k_2)$ that is the probability with which a random link is connected to a node with $k_1$ under the condition that the other end node has $k_2$.
For the random bimodal graph, we derive: 
\begin{align}
\langle k^n \rangle &= a k_1^n + (1-a) k_2^n \ (n = 1, 2,3, \cdots),\\
\overline{k}_{\rm nn}(k) &=
\begin{cases}
 k_1 \left\{ 1 - \frac{(1-a) k_2 P(k_1 | k_2)}{a k_1} \right\} + k_2 \frac{(1-a) k_2 P(k_1 | k_2)}{a k_1} & (k = k_1),\\
k_1 P(k_1 | k_2) + k_2 \left( 1 - P(k_1 | k_2) \right) & (k = k_2).
\end{cases}
\end{align}
When we substitute them into Eq.~\eqref{eq:r} and simplify the equation, we obtain
\begin{align}
r = 1 - \frac{(a k_1 + (1-a) k_2) P(k_1 | k_2)}{a k_1}.
\label{eq:r_bimodal}
\end{align}

\section*{Appendix B: Analytical expression of $\overline{\beta}$ and $\overline{\beta}_{\rm MFC}$ for the random bimodal graph}
\setcounter{equation}{0}
\renewcommand{\theequation}{B\arabic{equation}}

For the random bimodal graph, $\overline{\beta}$ (Eq.~\eqref{eq:sufficient_richclub}) is equal to the smaller one in the two terms, i.e, either of the cases with $k_c = k_1$ and $k_c = k_2$.
We obtain $E_c$ for $k_c = k_1$ and $k_c = k_2$ as
\begin{align}
E_c &=
\begin{cases}
M P(k_1, k_1) = M \frac{k_1 P(k_1 | k_1)}{\langle k \rangle} P(k_1) = \frac{a N k_1}{2} \left\{ 1 - \frac{(1-a) k_2 P(k_1 | k_2)}{a k_1}\right\} & (k_c = k_1),\\
M & (k_c = k_2),\\
\end{cases}
\label{eq:E_c}
\end{align}
where $M$ is the total number of links and we use identity $M = \langle k \rangle N/2$.
In addition, $K_c$ is given by 
\begin{align}
K_c =
\begin{cases}
N k_1 p(k_1) = a N k_1 & (k_c = k_1), \\
2 M  & (k_c = k_2). 
\end{cases}
\label{eq:K_c}
\end{align}
By combining Eqs.~\eqref{eq:E_c} and \eqref{eq:K_c}, we derive
\begin{align}
1 - \frac{2 E_c}{ K_c} =
\begin{cases}
\frac{(1-a)k_2 P(k_1 | k_2)}{a k_1} = (1-r) \frac{(1-a)k_2}{\langle k \rangle} & (k_c = k_1),\\
0 & (k_c = k_2).
\end{cases}
\label{eq:1-2EC}
\end{align}
To derive the equation for the case with $k_c = k_1$, we use
\begin{align}
P(k_1 | k_2) = (1-r) \frac{a k_1}{\langle k \rangle},
\label{eq:p_k1_k2}
\end{align}
which is obtained by transforming Eq.~\eqref{eq:r_bimodal}.
Finally, we derive
\begin{align}
\overline{\beta} = {\rm min}  \left[ \left( \mu + (1-r) \frac{(1-a)k_2}{\langle k \rangle} \Di \right) \left( \frac{k_1}{\langle k \rangle} \rho \right)^{-\gamma}, \mu \left( \frac{k_2}{\langle k \rangle} \rho \right)^{-\gamma} \right].
\label{eq:beta_bimodal}
\end{align}
As a reference, $\overline{\beta}_{\rm MFC}$ (Eq.~\eqref{eq:sufficient_mfc}) for the random bimodal graph is derived as
\begin{eqnarray}
\overline{\beta}_{\rm MFC} = {\rm min}  \left[ \left( \mu + (1-r) \frac{(1-a)k_2}{\langle k \rangle} \Di \right) \left( \frac{k_1}{\langle k \rangle} \rho \right)^{-\gamma},
\left( \mu + (1-r) \frac{a k_1}{\langle k \rangle} \Di \right) \left( \frac{k_2}{\langle k \rangle} \rho \right)^{-\gamma} \right],
\label{eq:mfc_bimodal}
\end{eqnarray}
where we use
\begin{align}
P(k_1 | k_1) &= 1 - P(k_2 | k_1) = 1- \frac{(1-a) k_2 P(k_1 | k_2)}{a k_1},\\
P(k_2 | k_2) &= 1-P(k_1 | k_2),
\end{align}
and Eq.~\eqref{eq:p_k1_k2}.
In Fig.~\ref{fig:bimodal_theory}, $\overline{\beta}$ and $\overline{\beta}_{\rm MFC}$ are plotted as a function of degree assortativity coefficient $r$ while the parameters are set to the values we used in Sec.~\ref{sec:bimodal}. For $\gamma = 0.1$ and $-1 \leq r \lesssim 0.5$, $\overline{\beta} < \overline{\beta}_{\rm MFC}$ holds. Otherwise, the two upper bounds are equivalent under these parameter settings.


\clearpage

\clearpage

\clearpage

\clearpage

\setcounter{figure}{0}
\renewcommand{\thefigure}{B\arabic{figure}}
\begin{figure}
\centering
\includegraphics[width=0.45\hsize]{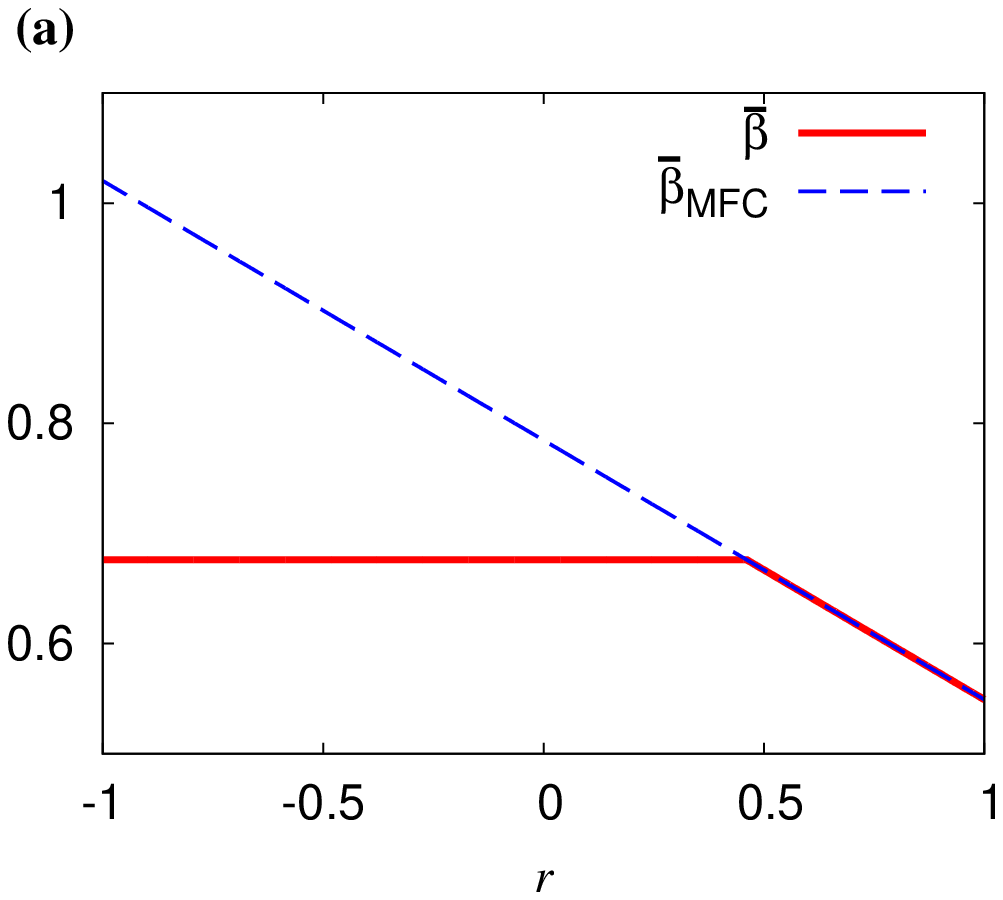}
\includegraphics[width=0.45\hsize]{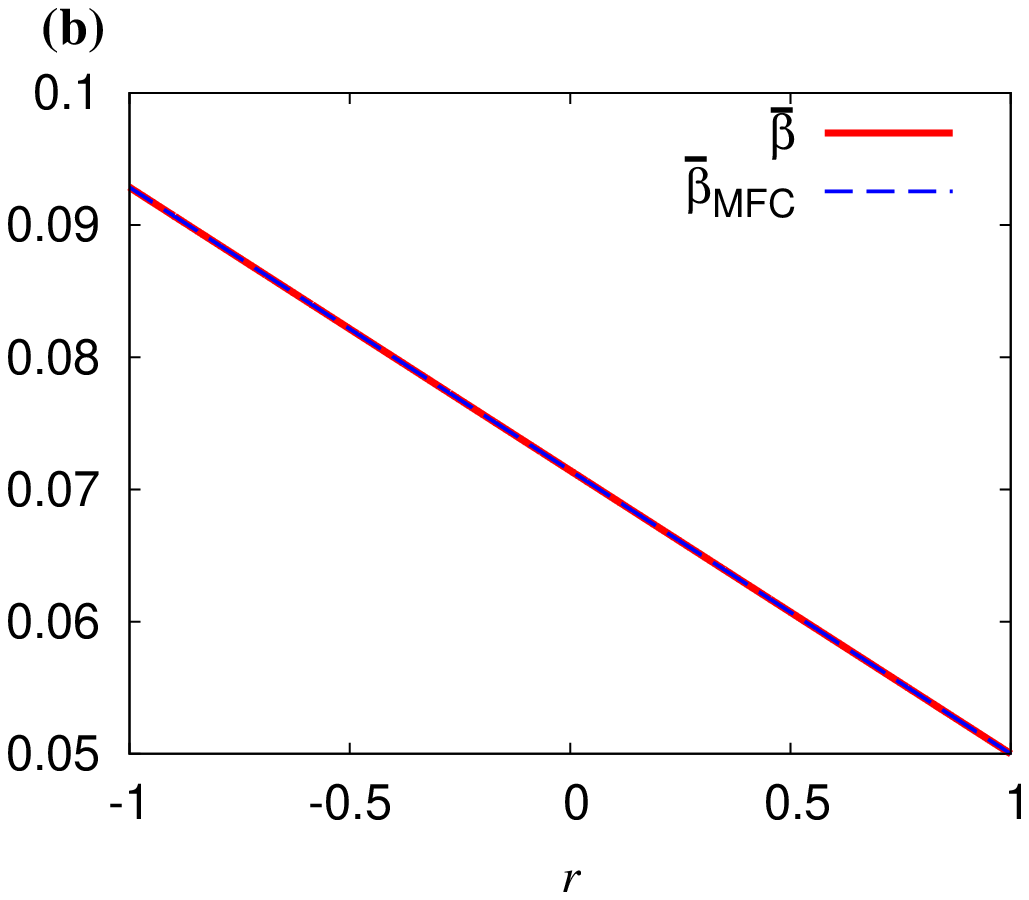}
\caption{
Upper bounds $\overline{\beta}$ (Eq.~\eqref{eq:beta_bimodal}, solid) and $\overline{\beta}_{\rm MFC}$ (Eq.~\eqref{eq:mfc_bimodal}, dashed) as a function of degree assortativity coefficient $r$ for the random bimodal graph. We set (a) $\gamma = 0.1$ and (b) $\gamma = 0.5$. Other parameters are fixed to $(k_1, k_2, a, \rho, \Di) = (16, 2, 100, 1)$, the same values as we used for the numerical simulations described in Sec.~\ref{sec:bimodal}.
}
\label{fig:bimodal_theory}
\end{figure}


\begin{thebibliography}{10}

\bibitem{Anderson1991}
Anderson,~R.~M. and May,~R.~M. 1991
\newblock {\em {Infectious diseases of humans: Dynamics and control}}.
\newblock Oxford University Press; Oxford.

\bibitem{Hanski2000}
Hanski,~I. and Ovaskainen,~O. 2000
\newblock {The metapopulation capacity of a fragmented landscape.}
\newblock {\em Nature} {\bf 404}, 755--758.
(doi:10.1038/35008063)

\bibitem{Colizza2007a}
Colizza,~V., Pastor-Satorras,~R., and Vespignani,~A. 2007
\newblock {Reaction-diffusion processes and metapopulation models in heterogeneous networks}.
\newblock {\em Nat. Phys.} {\bf 3}, 276--282.
(doi:10.1038/nphys560)

\bibitem{Zipf1949}
Zipf,~G.~K. 1949
\newblock {\em {Human behavior and the principle of least effort: An
  introduction to human ecology}}.
\newblock Addison-Wesley Press; Oxford.

\bibitem{Gabaix1999}
Gabaix,~X. 1999
\newblock {Zipf's law for cities: an explanation}.
\newblock {\em Q. J. Econ.} {\bf 114}, 739--767.
(doi:10.1162/003355399556133)

\bibitem{Carrothers1956}
Carrothers,~G.~A.~P. 1956
\newblock {An historical bedew of the gravity and potential concepts of human
  interaction}.
\newblock {\em J. Am. Inst. Plann.} {\bf 22}, 94--102.
(doi:10.1080/01944365608979229)

\bibitem{Saldana2008}
Salda\~{n}a,~J. 2008
\newblock {Continuous-time formulation of reaction-diffusion processes on
  heterogeneous metapopulations}.
\newblock {\em Phys. Rev. E} {\bf 78}, 012902.
(doi:10.1103/PhysRevE.78.012902)

\bibitem{Colizza2008}
Colizza,~V. and Vespignani,~A. 2008
\newblock {Epidemic modeling in metapopulation systems with heterogeneous coupling pattern: theory and simulations.}
\newblock {\em J. Theor. Biol.} {\bf 251}, 450--67.
(doi:10.1016/j.jtbi.2007.11.028)

\bibitem{Juher2009}
Juher,~D., Ripoll,~J., and Salda\~{n}a~J. 2009
\newblock {Analysis and Monte Carlo simulations of a model for the spread of infectious diseases in heterogeneous metapopulations}.
\newblock {\em Phys. Rev. E} {\bf 80}, 041920.
(doi:10.1103/PhysRevE.80.041920)

\bibitem{Saldana2010}
Salda\~{n}a~J. 2010
\newblock {Modelling the spread of infectious diseases in complex metapopulations}.
\newblock {\em Math. Model. Nat.} {\bf 5}, 22--37.
(doi:10.1051/mmnp/20105602)

\bibitem{Juher2014}
Juher,~D. and Ma\~{n}osa,~V. 2014
\newblock {Spectral properties of the connectivity matrix and the SIS-epidemic
  threshold for mid-size metapopulations}.
\newblock {\em Math. Model. Nat.} {\bf 9}, 108--120.
(doi:10.1051/mmnp/20149207)

\bibitem{Smith2009}
Smith,~M.~J., Telfer,~S., Kallio,~E.~R., Burthe,~S., Cook,~A.~R., Lambin,~X., and Begon,~M. 2009
\newblock {Host-pathogen time series data in wildlife support a transmission
  function between density and frequency dependence.}
\newblock {\em Proc. Natl. Acad. Sci. U.S.A.} {\bf 106}, 7905--7909.
(doi:10.1073/pnas.0809145106)

\bibitem{Earn2000}
Earn,~D.~J. 2000
\newblock {A simple model for complex dynamical transitions in epidemics}.
\newblock {\em Science} {\bf 287}, 667--670.
(doi:10.1126/science.287.5453.667)

\bibitem{Tildesley2006}
Tildesley,~M.~J., Savill,~N.~J., Shaw,~D.~J.  Deardon,~R., Brooks,~S.~P., Woolhouse,~M.~E~J. Grenfell,~B.~T., and Keeling,~M.~J. 2006
\newblock {Optimal reactive vaccination strategies for a foot-and-mouth outbreak in the UK.}
\newblock {\em Nature} {\bf 440}, 83--86.
(doi:10.1038/nature04324)

\bibitem{May1987}
May,~R.~M. and Anderson,~R.~M. 1987
\newblock {Transmission dynamics of HIV infection.}
\newblock {\em Nature} {\bf 326}, 137--142.
(doi:10.1038/326137a0)

\bibitem{McCallum2001}
McCallum,~H., Barlow,~N., and Hone,~J. 2001
\newblock {How should pathogen transmission be modelled?}
\newblock {\em Trends Ecol. Evol.} {\bf 16}, 295--300.
(doi:10.1016/S0169-5347(01)02144-9)

\bibitem{Bettencourt2007}
Bettencourt,~L.~M.~A., Lobo,~J., Helbing,~D., K\"{u}hnert,~C., and West,~G.~B. 2007
\newblock {Growth, innovation, scaling, and the pace of life in cities.}
\newblock {\em Proc. Natl. Acad. Sci. U.S.A.} {\bf 104}, 7301--7306.
(doi:10.1073/pnas.0610172104)

\bibitem{Schlapfer2014}
Schl\"{a}pfer,~M., Bettencourt,~L.~M.~A., Grauwin,~S., Raschke,~M., Claxton,~R., Smoreda,~Z., West,~G.~B., and Ratti,~C. 2014
\newblock {The scaling of human interactions with city size}.
\newblock {\em J. R. Soc. Interface} {\bf 11}, 20130789.
(doi:10.1098/rsif.2013.0789)

\bibitem{Masuda2010}
Masuda,~N. 2010
\newblock {Effects of diffusion rates on epidemic spreads in metapopulation networks}.
\newblock {\em New J. Phys.} {\bf 12}, 093009.
(doi:10.1088/1367-2630/12/9/093009)

\bibitem{Iggidr2012}
Iggidr,~A., Sallet,~G., and Tsanou,~B. 2012
\newblock {Global stability analysis of a metapopulation SIS epidemic model}.
\newblock {\em Math. Popul. Stud.} {\bf 19}, 115--129.
(doi:10.1080/08898480.2012.693844)

\bibitem{Lund2013}
Lund,~H., Lizana,~L., and Simonsen,~I. 2013
\newblock {Effects of city-size heterogeneity on epidemic spreading in a
  metapopulation: a reaction-diffusion approach}.
\newblock {\em J. Stat. Phys.} {\bf 151}, 367--382.
(doi:10.1007/s10955-013-0690-3)

\bibitem{Dorogovtsev2008}
Dorogovtsev,~S.~N., Goltsev,~A.~V., and Mendes,~J.~F.~F. 2008
\newblock {Critical phenomena in complex networks}.
\newblock {\em Rev. Mod. Phys.} {\bf 80}, 1275--1335.
(doi:10.1103/RevModPhys.80.1275)

\bibitem{Anderson1982}
Anderson,~R.~M. and May,~R.~M. 1982
\newblock {Directly transmitted infections diseases: control by vaccination}.
\newblock {\em Science} {\bf 215}, 1053--1060.
(doi:10.1126/science.7063839)

\bibitem{Cohen2000}
Cohen,~R., Erez,~K., ben-Avraham,~D., and Havlin,~S. 2000
\newblock {Resilience of the Internet to random breakdowns}.
\newblock {\em Phys. Rev. Lett.} {\bf 85}, 4626--4628.
(doi:10.1103/PhysRevLett.85.4626)

\bibitem{Callaway2000}
Callaway,~D.~S., Newman,~M.~E.~J., Strogatz,~S.~H., and Watts,~D.~J. 2000
\newblock {Network robustness and fragility: percolation on random graphs.}
\newblock {\em Phys. Rev. Lett.} {\bf 85}, 5468--5471.
(doi:10.1103/PhysRevLett.85.5468)

\bibitem{Pastor-Satorras2001}
Pastor-Satorras,~R. and Vespignani,~A. 2001
\newblock {Epidemic spreading in scale-free networks}.
\newblock {\em Phys. Rev. Lett.} {\bf 86}, 3200--3203.
(doi:10.1103/PhysRevLett.86.3200)

\bibitem{Pastor-Satorras2002}
Pastor-Satorras,~R. and Vespignani,~A. 2002
\newblock {Immunization of complex networks}.
\newblock {\em Phys. Rev. E} {\bf 65}, 036104.
(doi:10.1103/PhysRevE.65.036104)

\bibitem{Chung1997}
Chung,~F.~R.~K. 1997
\newblock {\em {Spectral Graph Theory}}.
\newblock American Mathematical Society; Providence, RI, USA.

\bibitem{Zhou2004}
Zhou,~S. and Mondrag\'{o}n,~R.~J. 2004
\newblock {The rich-club phenomenon in the Internet topology}.
\newblock {\em IEEE Comm. Lett.} {\bf 8}, 180--182.
(doi:10.1109/LCOMM.2004.823426)

\bibitem{Colizza2006}
Colizza,~V., Flammini,~A., Serrano,~M.~A., and Vespignani,~A. 2006
\newblock {Detecting rich-club ordering in complex networks.}
\newblock {\em Nat. Phys.} {\bf 2}, 110--115.
(doi:10.1038/nphys209)

\bibitem{Dorogovtsev2006}
Dorogovtsev,~S.~N., Goltsev,~A.~V., and Mendes,~J.~F.~F. 2006
\newblock {k-Core organization of complex networks}.
\newblock {\em Phys. Rev. Lett.} {\bf 96}, 040601.
(doi:10.1103/PhysRevLett.96.040601)

\bibitem{Oliveira2005}
de~Oliveira,~M. and Dickman,~R. 2005
\newblock {How to simulate the quasistationary state}.
\newblock {\em Phys. Rev. E} {\bf 71}, 016129.
(doi:10.1103/PhysRevE.71.016129)

\bibitem{Ferreira2011}
Ferreira,~S.~C., Ferreira,~R.~S., and Pastor-Satorras,~R. 2011
\newblock {Quasistationary analysis of the contact process on annealed scale-free networks}.
\newblock {\em Phys. Rev. E} {\bf 83}, 066113.
(doi:10.1103/PhysRevE.83.066113)

\bibitem{Mata2013}
Mata,~A.~S., Ferreira,~S.~C., and Pastor-Satorras,~R. 2013
\newblock {Effects of local population structure in a reaction-diffusion model of a contact process on metapopulation networks}.
\newblock {\em Phys. Rev. E} {\bf 88}, 042820.
(doi:10.1103/PhysRevE.83.066113)

\bibitem{Barthelemy2010}
Barth\'{e}lemy,~M., Godr\`{e}che,~C., and Luck,~J.-M. 2010
\newblock {Fluctuation effects in metapopulation models: percolation and pandemic threshold.}
\newblock {\em J. Theor. Biol.} {\bf 267}, 554--564.
(doi:10.1016/j.jtbi.2010.09.015)

\bibitem{Molloy1995}
Molloy,~M. and Reed,~B. 1995
\newblock {A critical point for random graphs with a given degree sequence}.
\newblock {\em Rand. Struct. Algorithms} {\bf 6}, 161--179.
(doi:10.1002/rsa.3240060204)

\bibitem{Newman2010}
Newman,~M.~E.~J. 2010
\newblock {\em {Networks: an Introduction}}.
\newblock Oxford University Press; Oxford.

\bibitem{Newman2002}
Newman,~M.~E.~J. 2002
\newblock {Assortative mixing in networks}.
\newblock {\em Phys. Rev. Lett.} {\bf 89}, 208701.
(doi:10.1103/PhysRevLett.89.208701)

\bibitem{Newman2003}
Newman,~M.~E.~J. 2003
\newblock {Mixing patterns in networks}.
\newblock {\em Phys. Rev. E} {\bf 67}, 026126.
(doi:10.1103/PhysRevE.67.026126)

\bibitem{Hasegawa2013}
Hasegawa,~T., Takaguchi,~T., and Masuda,~N. 2013
\newblock {Observability transitions in correlated networks}.
\newblock {Phys. Rev. E} {\bf 88}, 042809.
(doi:10.1103/PhysRevE.88.042809)

\bibitem{Opsahl2011}
Opsahl,~T. 2011
\newblock {Why {A}nchorage is not (that) important: {B}inary ties and {S}ample selection}.
\newblock {\url{http://toreopsahl.com/2011/08/12/why-anchorage-is-not-that-important-binary-ties-and-sample-selection/}}

\bibitem{Kunegis2013}
Kunegis,~J. 2013
\newblock {KONECT: the Koblenz Network Collection}.
\newblock In {\em Proceedings of the 22nd International Conference on World
  Wide Web Companion}, pp. 1343--1350, Rio de Janeiro, Brazil, ACM.

\bibitem{Castellano2012}
Castellano,~C. and Pastor-Satorras,~R. 2012
\newblock {Competing activation mechanisms in epidemics on networks}.
\newblock {\em Sci. Rep.} {\bf 2}, 371.
(doi:10.1038/srep00371)

\bibitem{Kitsak2010}
Kitsak,~M., Gallos,~L.~K., Havlin,~S., Liljeros,~F., Muchnik,~L., Stanley,~H.~E., and Makse,~H.~A. 2010
\newblock {Identification of influential spreaders in complex networks}.
\newblock {\em Nat. Phys.} {\bf 5}, 888--893.
(doi:10.1038/nphys1746)

\bibitem{Laumann1999}
Laumann, E. O., Youm, Y. 1999
\newblock{Racial/ethnic group differences in the prevalence of sexually transmitted diseases in the United States: A network explanation}.
\newblock{Sex. Transm. Dis.} {\bf 26}, 250--261.
(doi:10.1097/00007435-199905000-00003)

\bibitem{Lowndes2002}
Lowndes, C. M., Alary, M., Meda, H., Gnintoungb\'{e}, C. A. B., Mukenge-Tshibaka, L., Adjovi, C., Buv\'{e}, A., Morison, L., Laourou, M., Kanhonou, L., and Anagonou, S. 2002
\newblock{Role of core and bridging groups in the transmission dynamics of HIV and STIs in Cotonou, Benin, West Africa}.
\newblock {Sex. Transm. Infect.} {\bf 78}, i69--i77.
(doi:10.1136/sti.78.suppl\_1.i69)

\bibitem{Jolly2002}
Jolly, A. M., and Wylie J. L. 2002
\newblock{Gonorrhoea and chlamydia core groups and sexual networks in Manitoba}.
\newblock{Sex. Transm. Infect.} {\bf 78}, i145--i151.
(doi:10.1136/sti.78.suppl\_1.i145)

\bibitem{Fortunato2010}
Fortunato, S. 2010
\newblock {Community detection in graphs}.
\newblock {\em Phys. Rep.} {\bf 486}, 75--174.
(doi:10.1016/j.physrep.2009.11.002)


\bibitem{Serrano2007}
Serrano M.A., Bogu\~{n}\'{a}, M., Pastor-Satorras, R., and Vespignani, A. 2007
\newblock {Correlations in Complex Networks}.
\newblock In: Caldarelli, G. and Vespignani, A. (eds.), {\em Large Scale Structure and Dynamics of Complex Networks: From Information Technology to Finance and Natural Science}, World Scientific, Singapore, pages 35--65 (chapter~3).

\end{thebibliography}
\end{document}